\documentclass[twocolumn]{aastex631}
\usepackage{xcolor}

\newcommand{\gaia}{{Gaia}}
\newcommand{\wise}{{WISE}}
\newcommand{\twomass}{{2MASS}}

\newcommand{\C}{\ensuremath{^{\circ}C\;}}

\shorttitle{Four Large Planets Orbiting TOI-4010}
\shortauthors{Kunimoto et al.}
\graphicspath{{./}{figures/}}
\begin{document}

\title{TOI-4010: A System of Three Large Short-Period Planets With a Massive Long-Period Companion}

\correspondingauthor{Michelle Kunimoto}
\email{mkuni@mit.edu}

\author[0000-0001-9269-8060]{Michelle Kunimoto}
\affiliation{Department of Physics and Kavli Institute for Astrophysics and Space Research, Massachusetts Institute of Technology,\\77 Massachusetts Avenue, Cambridge, MA 02139, USA}

\author[0000-0001-7246-5438]{Andrew Vanderburg}
\affiliation{Department of Physics and Kavli Institute for Astrophysics and Space Research, Massachusetts Institute of Technology,\\77 Massachusetts Avenue, Cambridge, MA 02139, USA}

\author[0000-0003-0918-7484]{Chelsea X. Huang}
\affiliation{Centre for Astrophysics, University of Southern Queensland, West Street,
Toowoomba, QLD 4350 Australia}

% Dynamical Analysis
\author[0000-0002-7451-4704]{M. Ryleigh Davis}
\affiliation{Division of Geological and Planetary Sciences, California Institute of Technology, Pasadena, CA 91125, USA}

% HARPS-N
\author[0000-0001-5600-3778]{Laura Affer}
\affiliation{INAF - Osservatorio Astronomico di Palermo, Piazza del Parlamento, 1, 90134, Palermo, PA, Italy}

% HARPS-N
\author[0000-0002-8863-7828]{Andrew Collier Cameron}
\affiliation{Centre for Exoplanet Science, SUPA School of Physics and Astronomy, University of St Andrews,\\North Haugh, St Andrews KY16 9SS, UK}

% HARPS-N
\author[0000-0002-9003-484X]{David Charbonneau}
\affiliation{Center for Astrophysics \textbar \ Harvard \& Smithsonian, 60 Garden Street, Cambridge, MA 02138, USA}

% HARPS-N
\author[0000-0003-1784-1431]{Rosario Cosentino}
\affiliation{Fundación Galileo Galilei - INAF, Rambla José Ana Fernandez Pérez 7, E-38712 Breña Baja, Tenerife, Spain}

% HARPS-N
\author{Mario Damasso}
\affiliation{INAF – Osservatorio Astrofisico di Torino, Via Osservatorio, 20, 10025 Pino Torinese TO, Italy}

% HARPS-N 
\author[0000-0002-9332-2011]{Xavier Dumusque}
\affiliation{Observatoire Astronomique de l'Universit\'e de Gen\`eve, Chemin Pegasi 51, 1290 Versoix, Switzerland}

% HARPS-N
\author[0000-0002-4272-4272]{A. F. Martínez Fiorenzano}
\affiliation{Fundación Galileo Galilei - INAF, Rambla José Ana Fernandez Pérez 7, E-38712 Breña Baja, Tenerife, Spain}

% HARPS-N 
\author[0000-0003-4702-5152]{Adriano Ghedina}
\affiliation{Fundación Galileo Galilei - INAF, Rambla José Ana Fernandez Pérez 7, E-38712 Breña Baja, Tenerife, Spain}

% HARPS-N
\author[0000-0001-9140-3574]{R. D. Haywood}
\altaffiliation{STFC Ernest Rutherford Fellow}
\affiliation{Astrophysics Group, University of Exeter, Exeter EX4 2QL, UK}

% HARPS-N
\author[0000-0003-4047-0771]{Florian Lienhard}
\affiliation{Astrophysics Group, Cavendish Laboratory, University of Cambridge, J.J. Thomson Avenue, Cambridge CB3 0HE, UK}

% HARPS-N
\author[0000-0003-3204-8183]{Mercedes L\'opez-Morales}
\affiliation{Center for Astrophysics \textbar \ Harvard \& Smithsonian, 60 Garden Street, Cambridge, MA 02138, USA}

% HARPS-N
\author[0000-0002-9352-5935]{Michel Mayor}
\affiliation{Observatoire Astronomique de l'Universit\'e de Gen\`eve, Chemin Pegasi 51, 1290 Versoix, Switzerland}

% HARPS-N
\author[0000-0002-9815-773X]{Francesco Pepe}
\affiliation{Observatoire Astronomique de l'Universit\'e de Gen\`eve, Chemin Pegasi 51, 1290 Versoix, Switzerland}

% HARPS-N
\author[0000-0002-4445-1845]{Matteo Pinamonti}
\affiliation{INAF – Osservatorio Astrofisico di Torino, Via Osservatorio, 20, 10025 Pino Torinese TO, Italy}

% HARPS-N
\author[0000-0003-1200-0473]{Ennio Poretti}
\affiliation{Fundación Galileo Galilei - INAF, Rambla José Ana Fernandez Pérez 7, E-38712 Breña Baja, Tenerife, Spain}

% HARPS-N
\author[0000-0002-2218-5689]{Jes\'us Maldonado}
\affiliation{INAF - Osservatorio Astronomico di Palermo, Piazza del Parlamento, 1, 90134, Palermo, PA, Italy}

% HARPS-N
\author[0000-0002-6379-9185]{Ken Rice}
\affiliation{SUPA, Institute for Astronomy, Royal Observatory, University of Edinburgh, Blackford Hill, Edinburgh EH93HJ, UK}
\affiliation{Centre for Exoplanet Science, University of Edinburgh, Edinburgh, UK}

% HARPS-N
\author[0000-0002-7504-365X]{Alessandro Sozzetti}
\affiliation{INAF – Osservatorio Astrofisico di Torino, Via Osservatorio, 20, 10025 Pino Torinese TO, Italy}

% HARPS-N
\author[0000-0001-8749-1962]{Thomas G. Wilson}
\affiliation{Centre for Exoplanet Science, SUPA School of Physics and Astronomy, University of St Andrews,\\North Haugh, St Andrews KY16 9SS, UK}

% HARPS-N
\author[0000-0001-7576-6236]{Stéphane Udry}
\affiliation{Observatoire Astronomique de l'Universit\'e de Gen\`eve, Chemin Pegasi 51, 1290 Versoix, Switzerland}

% Keck2 AO/NRM
\author[0000-0002-9306-1704]{Jay Baptista}
\affiliation{Institute for Astronomy, University of Hawai`i, 2680 Woodlawn Drive, Honolulu, HI 96822, USA}
\affiliation{Department of Astronomy and Astrophysics, Yale University, New Haven, CT 06520, USA}

%khalid.barkaoui@uliege.be   (LCO data analysis)
\author[0000-0003-1464-9276]{Khalid Barkaoui}
\affiliation{Astrobiology Research Unit, Universit\'e de Li\`ege, 19C All\'ee du 6 Ao\^ut, 4000 Li\`ege, Belgium}
\affiliation{Department of Earth, Atmospheric and Planetary Science, Massachusetts Institute of Technology, 77 Massachusetts Avenue, Cambridge, MA 02139, USA}
\affiliation{Instituto de Astrof\'isica de Canarias (IAC), Calle V\'ia L\'actea s/n, 38200, La Laguna, Tenerife, Spain}

\author[0000-0002-7733-4522]{Juliette Becker}
\affiliation{Division of Geological and Planetary Sciences, California Institute of Technology, Pasadena, CA 91125, USA}

%pbenni@verizon.net   (Acton Sky Portal observations and data analysis)
\author[0000-0001-6981-8722]{Paul Benni}
\affiliation{Acton Sky Portal private observatory, Acton, MA, USA}

%abieryla@cfa.harvard.edu  (KeplerCam observations and data analysis, TRES)
\author[0000-0001-6637-5401]{Allyson Bieryla}
\affiliation{Center for Astrophysics \textbar \ Harvard \& Smithsonian, 60 Garden Street, Cambridge, MA 02138, USA}

%paubosch1998@gmail.com  (OAA observations and analysis)
\author[0000-0002-1514-5558]{Pau Bosch-Cabot}
\affiliation{Observatori Astron\`omic Albany\`a, Camí de Bassegoda S/N, Albany\`a 17733, Girona, Spain}

% SG3
\author[0000-0002-5741-3047]{David R. Ciardi}
\affiliation{Caltech/IPAC-NExScI, M/S 100-22, 770 S Wilson Avenue, Pasadena, CA 91106 USA}

%karen.collins@cfa.harvard.edu  (SG1/LCO Observations coordination/LCO time/Analysis)
\author[0000-0001-6588-9574]{Karen A.\ Collins}
\affiliation{Center for Astrophysics \textbar \ Harvard \& Smithsonian, 60 Garden Street, Cambridge, MA 02138, USA}

% SG1
\author[0000-0003-2781-3207]{Kevin I.\ Collins}
\affiliation{George Mason University, 4400 University Drive, Fairfax, VA, 22030 USA}

% Keck2 NRM/AO
\author{Elise Evans}
\affiliation{SUPA, Institute for Astronomy, Royal Observatory, University of Edinburgh, Blackford Hill, Edinburgh EH93HJ, UK}
\affiliation{Centre for Exoplanet Science, University of Edinburgh, Edinburgh, UK}

\author[0000-0001-9823-1445]{Trent J.~Dupuy}
\affiliation{SUPA, Institute for Astronomy, Royal Observatory, University of Edinburgh, Blackford Hill, Edinburgh EH93HJ, UK}
\affiliation{Centre for Exoplanet Science, University of Edinburgh, Edinburgh, UK}
% SG3 SAI observations
\author[0000-0003-2228-7914]{Maria V. Goliguzova}
\affiliation{Sternberg Astronomical Institute Lomonosov Moscow State University \\ 119992, Universitetskij prospekt 13, Moscow, Russian Federation}

% SG1
\author[0000-0002-4308-2339]{Pere Guerra}
\affiliation{Observatori Astron\`omic Albany\`a, Camí de Bassegoda S/N, Albany\`a 17733, Girona, Spain}

\author[0000-0001-9811-568X]{Adam Kraus}
\affiliation{Department of Astronomy, The University of Texas at Austin, Austin, TX 78712, USA}

\author[0000-0001-6513-1659]{Jack J. Lissauer}
\affiliation{NASA Ames Research Center, Moffett Field, CA 94035, USA}

% Keck2 AO/NRM
\author[0000-0001-8832-4488]{Daniel Huber}
\affiliation{Institute for Astronomy, University of Hawai`i, 2680 Woodlawn Drive, Honolulu, HI 96822, USA}

% SG1
\author[0000-0001-9087-1245]{Felipe Murgas}
\affiliation{Instituto de Astrof\'\i sica de Canarias (IAC), 38205 La Laguna, Tenerife, Spain}
\affiliation{Departamento de Astrof\'\i sica, Universidad de La Laguna (ULL), 38206, La Laguna, Tenerife, Spain}

%epalle@iac.es     (LCO time contribution)
\author[0000-0003-0987-1593]{Enric Palle}
\affiliation{Instituto de Astrof\'\i sica de Canarias (IAC), 38205 La Laguna, Tenerife, Spain}
\affiliation{Departamento de Astrof\'\i sica, Universidad de La Laguna (ULL), 38206, La Laguna, Tenerife, Spain}

\author[0000-0002-8964-8377]{Samuel N. Quinn}
\affiliation{Center for Astrophysics \textbar \ Harvard \& Smithsonian, 60 Garden Street, Cambridge, MA 02138, USA}

%SG3 SAI observations
\author[0000-0003-1713-3208]{Boris S. Safonov}
\affiliation{Sternberg Astronomical Institute Lomonosov Moscow State University \\ 119992, Universitetskij prospekt 13, Moscow, Russian Federation}

%rpschwarz@comcast.net  (LCO data analysis)
\author[0000-0001-8227-1020]{Richard P. Schwarz}
\affiliation{Center for Astrophysics \textbar \ Harvard \& Smithsonian, 60 Garden Street, Cambridge, MA 02138, USA}

%LCO and TESS Contributing Author
\author[0000-0002-1836-3120]{Avi Shporer}
\affiliation{Department of Physics and Kavli Institute for Astrophysics and Space Research, Massachusetts Institute of Technology,\\77 Massachusetts Avenue, Cambridge, MA 02139, USA}

%SED analysis
\author[0000-0002-3481-9052]{Keivan G.\ Stassun}
\affiliation{Department of Physics and Astronomy, Vanderbilt University, Nashville, TN 37235, USA}

% Architect
\author[0000-0002-4715-9460]{Jon M. Jenkins}
\affiliation{NASA Ames Research Center, Moffett Field, CA 94035, USA}

\author[0000-0001-9911-7388]{David W. Latham}
\affiliation{Center for Astrophysics \textbar \ Harvard \& Smithsonian, 60 Garden Street, Cambridge, MA 02138, USA}

% Architect
\author[0000-0003-2058-6662]{George R. Ricker}
\affiliation{Department of Physics and Kavli Institute for Astrophysics and Space Research, Massachusetts Institute of Technology,\\77 Massachusetts Avenue, Cambridge, MA 02139, USA}

% Architect
\author[0000-0002-6892-6948]{Sara Seager}
\affiliation{Department of Earth, Atmospheric and Planetary Science, Massachusetts Institute of Technology, 77 Massachusetts Avenue, Cambridge, MA 02139, USA}
\affiliation{Department of Physics and Kavli Institute for Astrophysics and Space Research, Massachusetts Institute of Technology,\\77 Massachusetts Avenue, Cambridge, MA 02139, USA}
\affiliation{Department of Aeronautics and Astronautics, Massachusetts Institute of Technology, Cambridge, MA 02139, USA}

% Architect
\author[0000-0001-6763-6562]{Roland Vanderspek}
\affiliation{Department of Physics and Kavli Institute for Astrophysics and Space Research, Massachusetts Institute of Technology,\\77 Massachusetts Avenue, Cambridge, MA 02139, USA}

% Architect
\author[0000-0002-4265-047X]{Joshua Winn}
\affiliation{Department of Astrophysical Sciences, Princeton University, 4 Ivy Lane, Princeton, NJ 08540, USA}

% TESS Contributing authors
\author[0000-0002-2482-0180]{Zahra Essack}
\affiliation{Department of Earth, Atmospheric and Planetary Science, Massachusetts Institute of Technology, 77 Massachusetts Avenue, Cambridge, MA 02139, USA}
\affiliation{Department of Physics and Kavli Institute for Astrophysics and Space Research, Massachusetts Institute of Technology,\\77 Massachusetts Avenue, Cambridge, MA 02139, USA}

\author[0000-0002-7871-085X]{Hannah M. Lewis}
\affiliation{Space Telescope Science Institute, 3700 San Martin Drive, Baltimore, MD 21218, USA}

\author[0000-0003-4724-745X]{Mark E. Rose}
\affiliation{NASA Ames Research Center, Moffett Field, CA 94035, USA}




\begin{abstract}
We report the confirmation of three exoplanets transiting TOI-4010 (TIC-352682207), a metal-rich K dwarf observed by TESS in Sectors 24, 25, 52, and 58. We confirm these planets with HARPS-N radial velocity observations and measure their masses with $8 - 12\%$ precision. TOI-4010 b is a sub-Neptune ($P = 1.3$ days, $R_{p} = 3.02_{-0.08}^{+0.08}~R_{\oplus}$, $M_{p} = 11.00_{-1.27}^{+1.29}~M_{\oplus}$) in the hot Neptune desert, and is one of the few such planets with known companions. Meanwhile, TOI-4010 c ($P = 5.4$ days, $R_{p} = 5.93_{-0.12}^{+0.11}~R_{\oplus}$, $M_{p} = 20.31_{-2.11}^{+2.13}~M_{\oplus}$) and TOI-4010 d ($P = 14.7$ days, $R_{p} = 6.18_{-0.14}^{+0.15}~R_{\oplus}$, $M_{p} = 38.15_{-3.22}^{+3.27}~M_{\oplus}$) are similarly-sized sub-Saturns on short-period orbits. Radial velocity observations also reveal a super-Jupiter-mass companion called TOI-4010 e in a long-period, eccentric orbit ($P \sim 762$ days and $e \sim 0.26$ based on available observations). TOI-4010 is one of the few systems with multiple short-period sub-Saturns to be discovered so far.
\end{abstract}

\keywords{Exoplanet Systems (484) --- Exoplanets (498) --- Exoplanet Dynamics (490) --- Hot Neptunes (754) --- Transit Photometry (1709)}

\section{Introduction} \label{sec:intro}

Two major classes of planetary system architectures have arisen from the census of exoplanets close to their stars: hot Jupiters ($R_{p} > 8~R_{\oplus}$, $P < 10$ days), and Neptune-sized and smaller planets ($R_{p} < 4~R_{\oplus}$) in compact multi-planet systems (``compact multis''). Hot Jupiters tend to have no co-planar transiting companions \cite[e.g.][]{Steffen2012, Hord2021} with few exceptions (e.g. WASP-47, \citealt{Becker2015}; TOI-1130, \citealt{Huang2020b}), have a wide range of orbital eccentricities, and are more common around metal-rich stars \citep{Fischer2005}. Meanwhile, compact multis tend to involve small planets in nearly circular and co-planar orbits \cite[e.g.][]{Xie2016, VanEylen2019}, and typically lack massive companions in similarly packed orbits.

Super-Neptune to sub-Saturn-sized planets ($R_{p} = 4 - 8~R_{\oplus}$) do not fit cleanly into either of the above classes. The most massive sub-Saturns (up to $\sim80~M_{\oplus}$) are similar to hot Jupiters in that they typically orbit metal-rich stars and are the only detected planets in their systems, while the least massive sub-Saturn-sized planets (down to $\sim5~M_{\oplus}$) tend to exist in low-eccentricity multi-planet systems \citep{Petigura2017}. However, very rarely do such systems feature sub-Saturns close to their stars. Only 63 sub-Saturns with $P < 20$ days in multi-planet systems have been confirmed so far, compared to 1213 smaller planets in these orbits.\footnote{According to the Planetary Composite Data Table on the \citet{NEA}, accessed 2023 March 13} Furthermore, only five known systems feature multiple sub-Saturns with $P < 20$ days: Kepler-18 \citep{Cochran2011}, Kepler-223 \citep{Mills2016}, K2-19 \citep{Armstrong2015}, V1298 Tau \citep{David2019}, and TOI-942 \citep{Carleo2021}. V1298 Tau and TOI-942 are both young stars ($\sim20 - 50$ Myr old), meaning their planets could be precursors to the more typical compact systems of small exoplanets found by Kepler. 

In this work, we present the confirmation of TOI-4010: a sixth example of a multi-planet system with two short-period sub-Saturns, in addition to an inner planet in the hot Neptune desert. We discuss the photometry, high-resolution imaging, and spectroscopy observations that resulted in the discovery and confirmation of the system in \S\ref{sec:obs}. In \S\ref{sec:star} and \S\ref{sec:planets}, we describe our determination of the stellar and planet parameters, as well as the detection of a fourth planet in precise radial velocity (RV) observations. In \S\ref{sec:discussion} we place TOI-4010 in the context of other multi-planet systems and discuss the implications of our results.

\section{Observations}\label{sec:obs}

\subsection{Transiting Exoplanet Survey Satellite}\label{sec:tess}

NASA's Transiting Exoplanet Survey Satellite \cite[TESS;][]{Ricker2014} launched in April 2018 with the goal of searching for transiting exoplanets around nearby, bright stars. While TESS targets $\sim20,000$ high-interest stars every 27.4-day sector with 2-minute cadence observations, it also enables the detection of exoplanets around millions more stars by regularly recording images of its entire field of view. These Full-Frame Images (FFIs) were saved every 30 minutes for the TESS Prime Mission (2018 July -- 2020 July), every 10 minutes for the first Extended Mission (2020 July -- 2022 September), and every 200 seconds for the second Extended Mission (2022 September --).

\subsubsection{Initial TESS Detections of TOI-4010}

TOI-4010 (TIC-352682207) was first observed by TESS in Sectors 24 (Camera 4, CCD 4, 2020 April 16 -- May 13) and 25 (Camera 4, CCD 3, 2020 May 13 -- August 06) only in the FFIs. These FFIs were calibrated using the TESS Image CAlibrator \citep[TICA;][]{TICA}, and further processed by the Quick-Look Pipeline \citep[QLP;][]{Huang2020}. QLP searched the Sector 24 and 25 multi-sector light curve of TIC-352682207 using Vartools \citep{HartmanBakos2016}, an implementation of the Box-Least Squares (BLS) transit search algorithm \citet{Kovacs2002}. This search detected a planet candidate with a radius of $R_{p} = 6.1\pm0.5~R_{\oplus}$ at an orbital period of $P = 5.4136\pm0.0004$ days, transit epoch of $T_{0} = 2000.047\pm0.002$ TBJD (BJD - 2457000), and transit signal-to-noise ratio of S/N = 20.

However, at $T = 11.5$ mag, the host star was not bright enough for the signal to be further processed by the standard QLP vetting process, which is reserved for stars brighter than $T = 10.5$ mag \citep{Guerrero2021}. The signal was later vetted as part of an independent review of QLP search results for planets around faint stars \citep{Kunimoto2022}, which also uncovered a similarly sized candidate with $R_{p} = 6.4\pm0.6~R_{\oplus}$, $P = 14.687\pm0.003$ days, $T_{0} = 1985.976\pm0.003$ TBJD, and S/N = 17. These signals were alerted on the MIT TESS Alerts portal\footnote{\url{https://tev.mit.edu/data/}} on 2021 June 23 as TOI-4010.01 and TOI-4010.02, respectively. 

A third, smaller candidate with $R_{p} = 3.3\pm0.6~R_{\oplus}$, $P = 1.3480\pm0.0004$ days, $T_{0} = 2007.533\pm0.009$ TBJD, and S/N = 8 was later detected with both an independent BLS search using a custom FFI light curve extraction and a multi-planet search from QLP. This signal was alerted as TOI-4010.03 on 2021 July 06.

The target was also observed in Sectors 52 (Camera 4, CCD 3, 2022 May 18 -- June 13) and 58 (Camera 2, CCD 1, 2022 October 29 -- November 26) in both the FFIs and at 2-minute cadence. 2-minute cadence data products were processed by the Science Processing Operations Center \citep[SPOC;][]{SPOC} at the NASA Ames Research Center. All three transiting planets were re-detected by QLP and SPOC in Sector 52, and passed all diagnostic criteria in the SPOC Data Validation (DV) reports \citep{Twicken2018, Li2019} including even-odd transits comparison, eclipsing binary discrimination tests, tests against effects associated with scattered light, and tests against background eclipsing binaries.

\subsubsection{This Work}

Here, we use custom 30-minute FFI light curves from Sectors 24 and 25 and the SPOC 2-minute cadence light curves from Sectors 52 and 58. We chose SPOC data for these sectors in order to take advantage of the faster cadence, which helps better resolve transit shapes. We followed the procedure of \citet{Vanderburg2019} to produce FFI light curves. In summary, after downloading the pixels surrounding TOI-4010 using \texttt{TESScut} \citep{Brasseur2019}, we extracted light curves from a series of both circular and irregularly shaped photometric apertures \citep{Vanderburg2016}. We then removed systematics by decorrelating each light curve with the mean and standard deviations of TESS spacecraft quaternion time series. Decorrelation was performed via linear regression with iterative outlier exclusion. We simultaneously modeled stellar variability by including a basis spline in our linear regression with breakpoints every 1.5 days. In-transit datapoints were manually masked out from the detrending process, and an extrapolation of the trends was used to estimate corrections during transit. We selected the final FFI light curve from the aperture that minimized the photometric scatter of out-of-transit data.

From Sectors 52 and 58, we used the SPOC Pre-search Data Conditioning Simple Aperture Photometry \cite[PDCSAP;][]{Smith2012,Stumpe2012,Stumpe2014} light curve with all data points with a nonzero quality flag removed. We further removed low-frequency trends from the PDCSAP light curve using the biweight time-windowed slider implemented in the \texttt{wotan} Python package \citep{Hippke2019} with a window of 0.5 days, and masked all in-transit points to avoid distorting the transit shapes.

The final four-sector light curve contained 2391 flux measurements at 30 minute cadence and 35,293 measurements at 2 minute cadence, with a total observing baseline of 954 days. We measured the 1 hour precision of the light curve to be 513 -- 583 ppm depending on the sector, estimated as 1.48 times the median absolute deviation (MAD) of the light curve after binning to 1 hour bins.

While TESS light curves are the primary data product for planet search and vetting, analysis of the TESS pixels can also help assess planet candidacy. Because photometers measure all light from pixels within a specific aperture, the flux from a nearby eclipsing binary (NEB) can contaminate the aperture and cause transit-like events in the light curve. These cases can be identified using difference imaging, whereby the difference of averaged in- and out-of-transit pixel images is found \citep{Bryson2013}. The location of the strongest flux variation in the difference image should correspond to the transit source.

\begin{figure*}[ht]
    \centering
    \includegraphics[width=\textwidth]{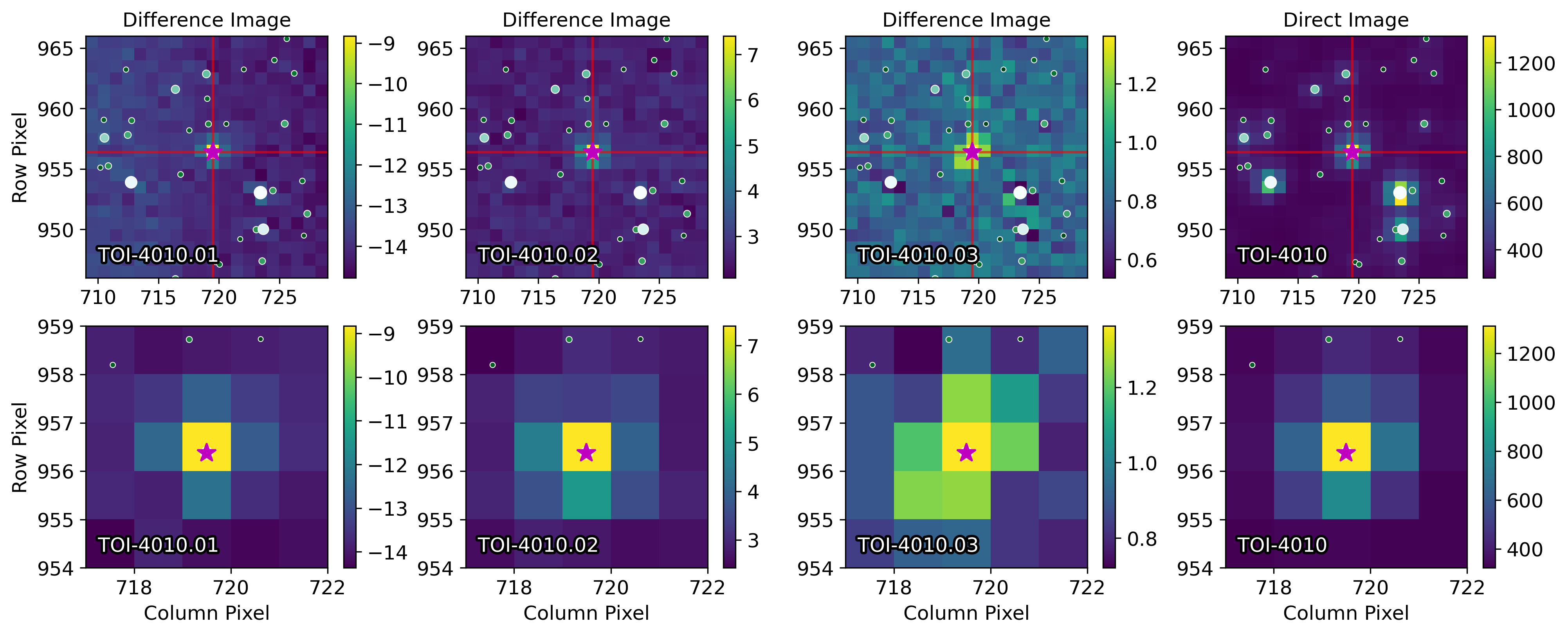}
    \caption{Top: Difference images for TOI-4010.01, 4010.02, and 4010.03, made using $20\times20$ pixel cutouts of the Sector 25 TESS FFIs. Bottom: A close-up of the central $5\times5$ pixels. The fourth column shows the average of out-of-transit images, which should represent a direct image of the field near the target star. The color bars show average pixel flux in electrons per second. The target TIC-352682207 is indicated with a pink star, while nearby stars up to $\Delta T = 4$ mag fainter are plotted as white circles with sizes scaled by brightness. The difference images for all three planet candidates indicate the transit sources are co-located with the target star.}
    \label{fig:diffimage}
\end{figure*}

To produce difference images, we used \texttt{TESS-plots}\footnote{\url{https://github.com/mkunimoto/TESS-plots}}, a Python package for difference image generation from TESS FFIs. For planets in multi-planet systems, \texttt{TESS-plots} masks out all cadences corresponding to other planet transits. This ensures that the primary source of variability in the difference image is due to the planet transit of interest. We inspected both Sector 24 and 25 images, with Sector 25 images shown in Figure \ref{fig:diffimage}. All difference images indicate that the three transit sources are co-located with the target star. These results are consistent with the difference images and centroid analysis from the Sector 52 SPOC DV reports, which find that the transit sources for all three planets are well within $3\sigma$ of the expected target location from the TESS Input Catalog \cite[TICv8.2;][]{Paegert2021}: the transit source for TOI-4010.01 is located within $1.7\pm2.6^{\prime\prime}$ of the target star, that for TOI-4010.02 is within $4.6\pm2.6^{\prime\prime}$, and TOI-4010.03 is within $1.9\pm2.8^{\prime\prime}$.

\subsection{Archival Images}

\begin{figure*}[t!]
    \centering
    \includegraphics[width=\linewidth]{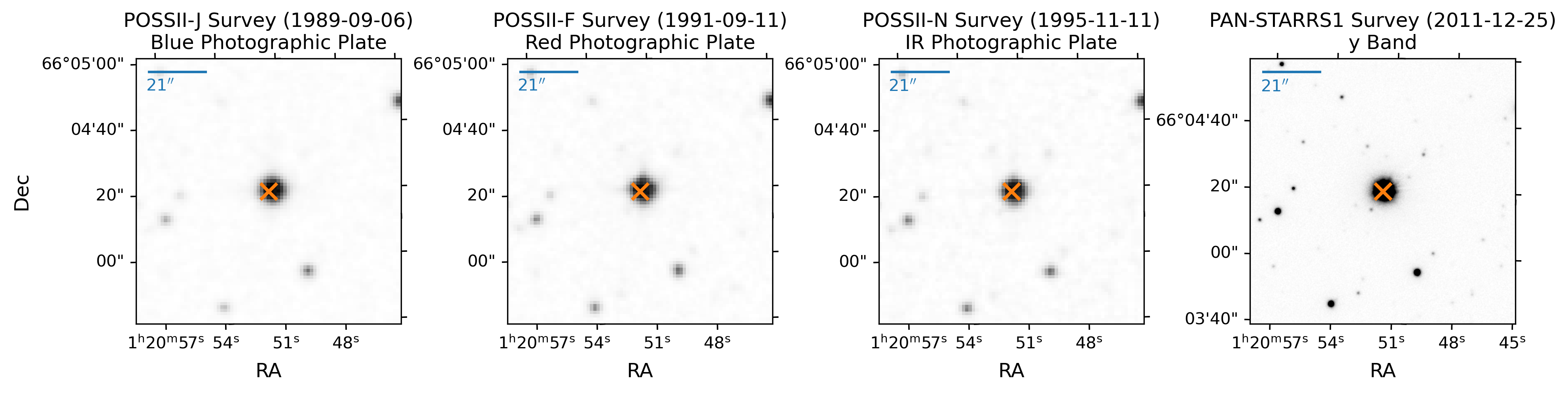}
    \caption{Archival images showing the field within $40^{\prime\prime}$ of TOI-4010 in blue (left), red (center-left), and infrared (center-right) filters from the second Palomar Sky Survey, and the $y$ band (right) from the first Pan-STARRS survey. The location of TOI-4010 at the start of TESS Sector 24 in 2020 is marked with an orange cross. TOI-4010 has moved only $\sim3^{\prime\prime}$ since the oldest image in 1989, and is relatively isolated, with no nearby brighter stars.}
    \label{fig:finders}
\end{figure*}

TOI-4010 was observed as part of the second Palomar Sky Survey \citep[POSS-II;][]{Reid1992} between 1989 and 1995 in blue, red, and infrared bands. We obtained images of the corresponding finder charts from the NASA/IPAC Infrared Science Archive,\footnote{\url{https://irsa.ipac.caltech.edu/applications/finderchart}} as shown in Figure \ref{fig:finders}. The star was also observed in the first survey by the Panoramic Survey Telescope and Rapid Response System \cite[Pan-STARRS1;][]{PanSTARRS1}. We obtained processed images in $g$, $r$, $i$, $z$, and $y$ from the Pan-STARRS1 image cutout service.\footnote{\url{http://ps1images.stsci.edu/cgi-bin/ps1cutouts}} Only the $y$ band cutout from 2011 is included in Figure \ref{fig:finders} due to saturation in other bands.

TOI-4010 has moved only $\sim3^{\prime\prime}$ since the oldest image in 1989, and is relatively isolated. All six \gaia\ sources within $21^{\prime\prime}$ (the size of one TESS pixel) are fainter by $\Delta T \approx 7$ mag or more, and are therefore too faint to reproduce the transits.

\subsection{Ground-Based Photometry}\label{sec:sg1}

We obtained ground-based photometric observations through the TESS Follow-up Observing Program (TFOP) SG1, which is specialized in seeing-limited photometry to aid in the validation of TESS planet candidates. Ground-based photometry can identify false positives due to NEBs at smaller separations than is achievable with difference image analysis. Observations can also provide improved transit ephemerides, independent transit depth constraints, and confirmation that the event is on-target. Our observations are summarized in Table \ref{tab:sg1}, and all time series are available on the Exoplanet Follow-up Observing Program website\footnote{\url{https://exofop.ipac.caltech.edu}} \cite[ExoFOP;][]{ExoFOP}. At least two on-target full transits of each planet were detected.

\begin{table*}[]
    \centering
    \caption{Summary of ground-based time series observations.}
    \begin{tabular}{cccccccl}
    \hline\hline
    TOI & Telescope & Date & Aperture & Filter & Observing & Number of & Notes \\
    & & (UTC) & ($^{\prime\prime}$) & & Duration (min) & Observations & \\
    \hline
    TOI-4010.01 & LCO-McD & 2021-08-28 & 3.90 & $gp$, $ip$ & 337 & 311 & Full transit \\
    & ASP & 2021-09-08 & 6.90 & $R_{c}$ & 420 & 903 & Full transit; poor weather\\
    & OAA & 2021-12-30 & 14.4 & $I_{c}$ & 669 & 303 & Full transit \\
    \hline
    TOI-4010.02 & LCO-Teide & 2021-08-30 & 5.85 & $gp$, $ip$ & 263 & 244 & Full transit \\
    & LCO-Teide & 2021-10-27 & 4.27 & $gp$, $ip$ & 451 & 416 & Full transit \\
    \hline
    TOI-4010.03 & LCO-McD & 2021-09-08 & 3.51 & $ip$ & 118 & 231 & No detection \\
    & FLWO & 2021-09-12 & 3.36 & $ip$ & 268 & 459 & No detection \\
    & LCO-Teide & 2021-10-03 & 3.51 & $ip$ & 381 & 437 & Ingress \\
    & LCO-Teide & 2021-10-08 & 5.46 & $ip$ & 342 & 379 & Ingress \\
    & LCO-McD & 2021-10-16 & 5.46 & $ip$ & 385 & 423 & Inconclusive \\
    & LCO-Teide & 2021-10-28 & 3.12 & $ip$ & 265 & 304 & Inconclusive \\
    & LCO-HAL & 2022-07-11 & 4.2 & $gp$, $ip$, $rp$, $zp$ & 181 & 1988 & Inconclusive \\
    & LCO-Teide & 2022-08-06 & 2.7 & $ip$ & 207 & 238 & Full transit \\
    & LCO-HAL & 2022-08-11 & 4.0 & $gp$, $ip$, $rp$, $zp$ & 218 & 2422 & Full transit; poor weather \\
    & LCO-Teide & 2022-08-24 & 5.1 & $ip$ & 201 & 196 & Full transit \\
    & LCO-Teide & 2022-09-06 & 3.5 & $ip$ & 208 & 209 & Full transit \\
    \\
    \end{tabular}
    \label{tab:sg1}
\end{table*}

\subsubsection{Las Cumbres Observatory}
Transits of all three planets were observed with telescopes in the Las Cumbres Observatory global telescope network \citep[LCOGT; ][]{Brown2013}. Observations taken by the 1m telescopes at the McDonald Observatory (LCO-McD) and Teide Observatory (LCO-Teide) use Sinistro cameras with a pixel scale of $0.389^{\prime\prime}$px$^{-1}$. Observations taken by the 2m telescope at the Haleakala Observatory (LCO-HAL) use a MuSCAT3 camera \citep{Narita2020} with a pixel scale of $0.265^{\prime\prime}$px$^{-1}$. Images were calibrated using the standard LCOGT \texttt{BANZAI} pipeline \citep{McCully2018}, and photometric data were extracted with \texttt{AstroImageJ} \citep{Collins2017}.

{\bf TOI-4010.01:} A full transit of TOI-4010.01 was observed on 2021 August 28 by LCO-McD in $gp$ and $ip$ bands, each giving depths and durations consistent with the TESS transits. A 10.0 px ($3.90^{\prime\prime}$) aperture optimized the photometric precision in both bands, and a smaller 5.0 px ($1.95^{\prime\prime}$) aperture was used to clear the field of NEBs at the transit ephemerides out to $2.5^{\prime}$.

{\bf TOI-4010.02:} A full transit of TOI-4010.02 was observed on 2021 August 30 by LCO-Teide in $gp$ and $ip$ bands. A 15.0 px ($5.85^{\prime\prime}$) aperture optimized the photometric precision, and a 5.0 px aperture was used to clear the field out to $2.5^{\prime}$. A second full transit was observed on 2021 October 27, optimized by an 11.0 px ($4.29^{\prime\prime}$) aperture. Both transits had depths and durations consistent with the TESS transits.
 
{\bf TOI-4010.03:} A first attempt to detect TOI-4010.03 was made on 2021 September 8 at LCO-McD. No transit on-target was observed, which was later attributed to the TESS ephemerides predicting the transit too early. Observations on 2021 October 03 at LCO-Teide tentatively detected the event in a 9.0 px ($3.51^{\prime\prime}$) aperture, but clarity depended on the choice of aperture size. TOI-4010.03 was also tentatively detected on October 08 at LCO-Teide and October 16 at LCO-McD in 14.0 px ($5.46^{\prime\prime}$) apertures, but a smaller 10.0 px aperture did not produce valid light curves. A further attempt on 2021 October 28 at LCO-Teide in the $ip$ band was inconclusive.

The star was later re-observed by TESS in Sector 52, allowing us to precisely constrain the ephemerides and revisit the LCO data. We determined that partial transits of TOI-4010.03 occurred in the LCO observations from October 03 and 08, and an egress was caught shortly before the transit of TOI-4010.02 on August 30. 

Post-TESS Sector 52, further attempts to detect TOI-4010.03 were made at LCO-HAL on 2022 July 11 and August 11 in $gp$, $rp$, $ip$, and $zs$ bands. Both attempts were affected by poor weather, though the latter recovered on-time transits in each band. Three on-time full transits were finally detected by LCO-Teide on 2022 August 06, August 24, and September 06, providing definitive ground-based detections of TOI-4010.03.

\subsubsection{Acton Sky Portal Observatory}

A full transit of TOI-4010.01 was observed on 2021 September 08 at the private observatory Acton Sky Portal (ASP), using an SBIG ST-8-XME camera mounted on a 0.36m Celestron EdgeHD 1400. The camera has a pixel scale of $0.69^{\prime\prime}$px$^{-1}$, and observations were taken with an $R_{c}$ filter. Cloud disruption affected data near ingress, and there is a systematic noise feature in the light curve due to an image reset. However, the transit depth and egress time matched predictions from the TESS observations, further supporting that TOI-4010.01 is on-target.

\subsubsection{Fred Lawrence Whipple Observatory}

A detection of TOI-4010.03 in the $ip$ band was attempted with the 1.2m telescope at the Fred Lawrence Whipple Observatory (FLWO) on 2021 September 12. The telescope uses the KeplerCam, which has a pixel scale of $0.672^{\prime\prime}$px$^{-1}$. We subsequently found that the observations were too early to cover the transit. Nevertheless, these were useful in determining that future observations should attempt to detect TOI-4010.03 later than predicted by the TESS data.

\subsubsection{Observatori Astron\`{o}mic Albany\`{a}}

A full transit of TOI-4010.01 was observed on 2021 December 30 at the Observatori Astron\`{o}mic Albany\`{a} using a Moravian G4-9000 camera mounted on the 0.4m OAA telescope. The camera has a pixel scale of $1.44^{\prime\prime}$px$^{-1}$ and observations were taken with an $I_{c}$ filter. Photometric data were extracted with \texttt{AstroImageJ}. The transit was detected on-target at the expected ephemeris, with the expected transit depth and duration, in an uncontaminated 10.0 px ($14.4^{\prime\prime}$) aperture.

\subsection{High-Resolution Imaging}

We obtained high-resolution imaging to detect nearby objects that are not resolved in the TIC or by seeing-limited photometry. Unresolved eclipsing binaries may be the source of a false positive signal, while light from nearby stars can also dilute the transit depth, thus reducing the inferred planet radius.

\subsubsection{W. M. Keck Observatory}

\begin{figure}[t]
    \centering
    \includegraphics[width=\linewidth]{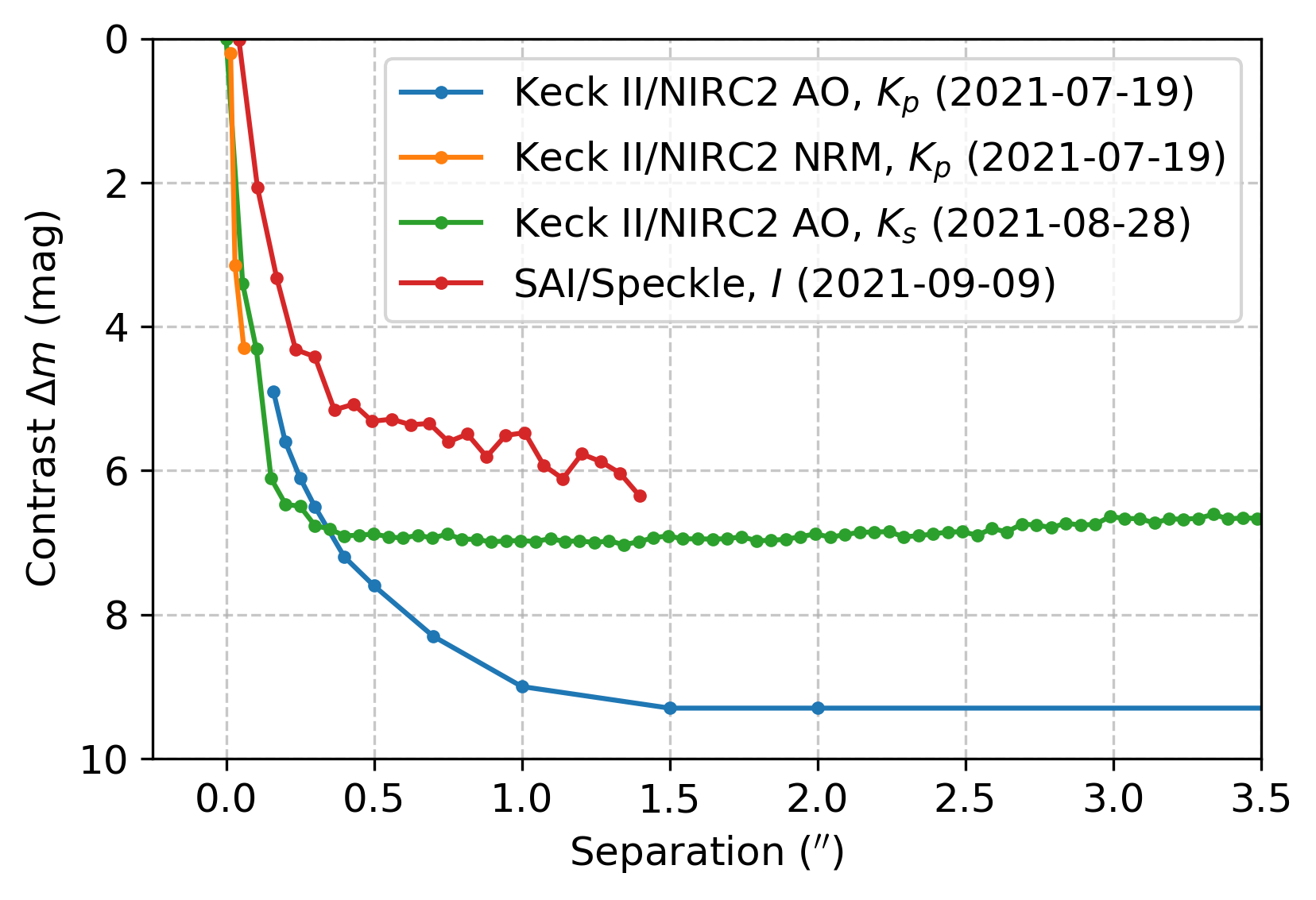}
    \caption{Contrast curves from Keck II/NIRC2 AO analysis in the $K_p$ band (blue), Keck II/NIRC2 NRM analysis in the $K_p$ band (orange), Keck II/NIRC2 AO analysis in the $K_s$ band (green), and SAI/Speckle analysis in the $I$ band (red). No companions bright enough to contaminate the TESS photometry or cause a false positive were detected.}
    \label{fig:highres}
\end{figure}

TOI-4010 was observed on 2021 July 19 by the 10m Keck II telescope at the W. M. Keck Observatory. We obtained adaptive optics (AO) imaging in the $K_p$ band with the Near-Infrared Camera (NIRC2),\footnote{\url{https://www2.keck.hawaii.edu/inst/nirc2/}} as well as data for non-redundant aperture masking interferometry \cite[NRM;][]{Tuthill2000}. The image processing and analysis was conducting following the methods described by \citet{Kraus2016}. Contrast curves from both types of observations are included in Figure \ref{fig:highres}. The AO images revealed a likely neighbor at a separation of $3.174\pm0.004^{\prime\prime}$, a position angle of $73.09\pm0.07^{\circ}$, and a contrast of $\Delta K_p = 8.58\pm0.07$ mag, placing it roughly 0.6 mag above the detection limits. This neighbor is too faint to reproduce the transits of TOI-4010, and thus we do not consider it a false positive source or significant contaminant to the light curve. No other neighbors were revealed in either the AO or NRM analysis.

We obtained additional NIRC2 AO contrast curves through TFOP SG3 based on observations from 2021 August 28, shown in Figure \ref{fig:highres}, demonstrating that there are no nearby stars or companions in the field of view down to the instrumental detection limits. Observations were made in the $K_s$ band with an estimated PSF of $0.05^{\prime\prime}$. These results are available on ExoFOP.

\subsubsection{Caucasian Observatory}

TOI-4010 was also observed on 2021 September 09 with the speckle polarimeter on the 2.5m telescope at the Caucasian Observatory of Sternberg Astronomical Institute (SAI) of Lomonosov Moscow State University \citep{Safonov2017}. Observations were made in the $I$ band with an estimated PSF of 0.083$^{\prime\prime}$ and obtained through TFOP SG3. The sensitivity curve is shown in Figure \ref{fig:highres}, again finding no detectable companions.

\subsection{High-Resolution Spectroscopy}\label{sec:rvs}

Spectroscopic observations enable precise characterization of planet host stars, and can identify false positives caused by spectroscopic binary stars. If sufficiently precise, RV data derived from spectroscopic observations can also further characterize orbits, constrain planet masses, and subsequently lead to the confirmation of the candidates as bonafide planets.

\subsubsection{TRES Spectroscopy}

We obtained a single reconnaissance spectrum of TOI-4010 on 2021 September 20 from the Tillinghast Reflector Echelle Spectrograph (TRES) through TFOP SG2. TRES has a spectral range from $385 - 910$ nm at a resolution of $R \sim 44,000$ over 51 orders \citep{Szentgyorgyi2007, Mink2011} and is located on the FLWO 1.5m Tillinghast Reflector telescope on Mount Hopkins, Arizona. The spectrum suggested that TOI-4010 was a slowly rotating K-dwarf suitable for more precise RV observations.

\subsubsection{HARPS-N Spectroscopy}

We obtained 112 spectra of TOI-4010 between 2021 July -- 2023 January with the High Accuracy Radial velocity Planet Searcher for the Northern Hemisphere \cite[HARPS-N;][]{Cosentino2012} at the 3.6m Telescopio Nazionale Galileo (TNG) at the Roque de los Muchachos Observatory. HARPS-N is a high resolution ($R = 115,000$) optical echelle spectrograph with a spectral range of $378 - 691$ nm, whose long-term pressure and temperature stability enable it to reach stability suitable for precise RV work. 

The spectra were reduced and had RVs extracted using the HARPS-N Data Reduction Software (DRS) v2.3.5. The RVs were determined by computing the center of a Gaussian fit to the cross correlation function (CCF) of the spectrum with a binary mask \citep{Baranne1996,Fellgett1955,Griffin1967}. The weight associated to each line in the mask is proportional to the RV content of the corresponding stellar line \citep{Dumusque2018,Pepe2002}. The resulting 112 RVs have a mean uncertainty of 5.0 m s$^{-1}$ and cover a total baseline of 545 days. 

\section{Stellar Parameters}\label{sec:star}

TIC-352682207 is listed in the TICv8.2 with $R_{\star} = 0.88\pm0.06~R_{\odot}$, $M_{\star} = 0.80\pm0.10~M_{\odot}$, $T_{\rm eff} = 4888\pm138$ K, and $\log g~\text{(cgs)}= 4.45\pm0.09$, implying a K dwarf. We further characterized the stellar properties using three different methods. A summary of stellar properties and their sources is given in Table \ref{tab:star}.

\begin{table*}[h]
    \centering
    \caption{Literature and Derived Parameters for TIC-352682207 (TOI-4010)}
    \begin{tabular}{l|l|l|l}
        \hline\hline
        Parameter & Value & Description & Source \\ 
        \hline\hline
        \multicolumn{4}{c}{\textit{TIC Parameters}} \\
        \hline
        ID & 352682207 & TESS Input Catalog ID & TICv8.2 \\
        $T_{\rm eff}$ & $4888\pm138$ & Effective temperature (K) & TICv8.2\\
        $\log g$ & $4.45\pm0.09$ & Surface gravity (cgs) & TICv8.2\\
        $R_{\star}$ & $0.88\pm0.06$ & Stellar radius ($R_{\odot}$) & TICv8.2 \\
        $M_{\star}$ & $0.80\pm0.10$ & Stellar mass ($M_{\odot}$) & TICv8.2 \\
        \hline
        \multicolumn{4}{c}{\textit{Astrometric Parameters}} \\
        \hline
        $\alpha$ & 01:20:51.50 & Right ascension (J2016) & \gaia\ DR3 \\
        $\delta$ & +66:04:19.91 & Declination (J2016) & \gaia\ DR3 \\
        $\bar{\omega}$ & $5.6646\pm0.0110$ & Parallax (mas) & \gaia\ DR3\ \\
        $\mu_{\alpha}$ & $38.4844\pm0.0088$& Proper motion right ascension (mas~yr$^{-1}$) & \gaia\ DR3 \\
        $\mu_{\delta}$ & $-20.5562\pm0.0116$ & Proper motion declination (mas~yr$^{-1}$) &\gaia\ DR3 \\
        \hline
        \multicolumn{4}{c}{\textit{Photometric Parameters}} \\
        \hline
        $T$ & $11.5163\pm0.0061$ & TESS band magnitude (mag) & TICv8.2 \\
        \hline
        $B$ & $13.381\pm0.026$ & $B$ band magnitude (mag) & UCAC4\\
        $V$ & $12.291\pm0.069$ & $V$ band magnitude (mag) & UCAC4\\
        $G$ & $12.1118\pm0.0028$ & $G$ band magnitude (mag) & \gaia\ DR3 \\
        $G_{BP}$ & $12.6217\pm0.0029$ & $G_{BP}$ band magnitude (mag) & \gaia\ DR3 \\
        $G_{RP}$ & $11.4534\pm0.0038$ & $G_{RP}$ band magnitude (mag) & \gaia\ DR3 \\
        $J$ & $10.649\pm0.027$ & $J$ band magnitude (mag) & \twomass \\
        $H$ & $10.25\pm0.031$ & $H$ band magnitude (mag) & \twomass \\
        $K_{s}$ & $10.157\pm0.021$ & $K$ band magnitude (mag) & \twomass \\
        $W1$ & $10.045\pm0.023$ & $W1$ band magnitude (mag) & \wise\\
        $W2$ & $10.137\pm0.021$ & $W2$ band magnitude (mag) & \wise \\
        $W3$ & $9.927\pm0.053$ & $W3$ band magnitude (mag) & \wise \\
        \hline
        \multicolumn{4}{c}{\textit{Spectroscopic Parameters}} \\
        \hline
        $T_{\rm eff}$ & $5104\pm50$ & Effective temperature (K) & TRES SPC analysis (\S\ref{sec:spc})\\
        $\log g$ & $4.584\pm0.1$ & Surface gravity (cgs) & TRES SPC analysis\\
        $[$Fe/H$]$ & $0.315\pm0.08$ & Metallicity (dex) & TRES SPC analysis\\
        $v\sin{i}$ & $3.31\pm0.5$ & Projected rotational velocity (km~s$^{-1}$)& TRES SPC analysis\\
        \hline
        $T_{\rm eff}$ & $4985\pm50$ & Effective temperature (K) & HARPS-N SPC analysis (\ref{sec:spc})\\
        $\log g$ & $4.450\pm0.1$ & Surface gravity (gs) & HARPS-N SPC analysis\\
        $[$Fe/H$]$ & $0.360\pm0.08$ & Metallicity (dex) & HARPS-N SPC analysis\\
        $v\sin{i}$ & $1.30\pm0.5$ & Projected rotational velocity (km~s$^{-1}$)& HARPS-N SPC analysis\\
        \hline
        \multicolumn{4}{c}{\textit{Derived Parameters}} \\
        \hline
        $F_{\rm bol}$ & $(4.09\pm0.14) \times 10^{-10}$ & Bolometric flux (erg~s$^{-1}$~cm$^{-2}$) & SED analysis (\S\ref{sec:sed}) \\
        $R_{\star}$ & $0.85\pm0.02$ & Stellar radius ($R_{\odot}$) & Empirical analysis \\
        $M_{\star}$ & $0.89\pm0.03$ & Stellar mass ($M_{\odot}$) & Empirical analysis \\
        $T_{\rm eff}$ & $4960\pm36$ & Effective temperature (K) & \texttt{isochrones} fit (\S\ref{sec:isochrones}) \\
        $\log g$ & $4.54\pm0.02$ & Surface gravity (cgs) & \texttt{isochrones} fit \\
        $[$Fe/H$]$ & $0.37\pm0.07$ & Metallicity (dex) & \texttt{isochrones} fit \\   
        $R_{\star}$ & $0.83\pm0.01$ & Stellar radius ($R_{\odot}$) & \texttt{isochrones} fit \\
        $M_{\star}$ & $0.88\pm0.03$ & Stellar mass ($M_{\odot}$) & \texttt{isochrones} fit \\
        $\tau_\star$ & $6.1\pm3.1$ & Stellar age (Gyr) & \texttt{isochrones} fit \\
        $\log{R^{\prime}_{HK}}$ & $-4.90\pm0.24$ & Calcium activity index & Stellar activity analysis (\S\ref{sec:activity}) \\
        $P_{\text{rot}}$ & $37.7\pm15.0$ & Stellar rotation period (days) & Stellar activity analysis \\
    \end{tabular}
    \label{tab:star}
\end{table*}

\subsection{Stellar Parameters Classification Tool}\label{sec:spc}

We used the Stellar Parameter Classification tool \cite[SPC;][]{Buchhave2012, Buchhave2014} to derive spectroscopic parameters from the TRES reconnaissance spectrum, obtaining $T_{\rm eff} = 5104\pm50$ K, $\log g~\text{(cgs)} = 4.584\pm0.1$, [Fe/H] $= 0.315\pm0.08$ dex, and $v\sin{i} = 3.31\pm0.5$ km~s$^{-1}$. When applied to the HARPS-N spectra, we find $T_{\rm eff} = 4985\pm50$ K, $\log g~\text{(cgs)} = 4.45\pm0.1$, [Fe/H] $= +0.36\pm0.08$ dex, and $v\sin{i} = 1.3\pm0.5$ km~s$^{-1}$. We adopt the HARPS-N SPC-derived values in later analysis (\S\ref{sec:sed}, \S\ref{sec:isochrones}) under the belief that they are more robust due to the significantly higher resolution and signal-to-noise ratio of the spectrograph.

\subsection{Spectral Energy Density Analysis}\label{sec:sed}

\begin{figure}[!ht]
\centering
\includegraphics[width=\linewidth,trim=75 70 80 70,clip]{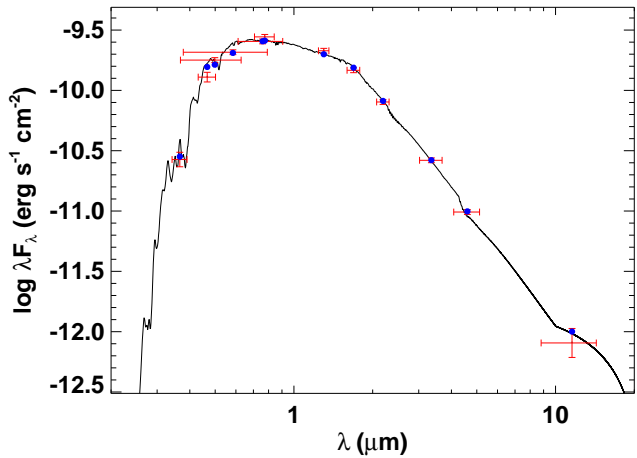}
\caption{Spectral energy distribution of TOI-4010. Red symbols represent the observed photometric measurements, where the horizontal bars represent the effective width of the passband. Blue symbols are the model fluxes from the best-fit Kurucz atmosphere model (black).  \label{fig:sed}}
\end{figure}

As an independent determination of the basic stellar parameters, we performed an analysis of the broadband spectral energy distribution (SED) of the star together with the \gaia\ DR3 parallax \citep{Gaia,GaiaDR3} \citep[with no systematic offset applied; see, e.g.,][]{StassunTorres:2021}, in order to determine an empirical measurement of the stellar radius, following the procedures described in \citet{Stassun:2016,Stassun:2017,Stassun:2018}. We pulled the $UBV$ magnitudes from \citet{Mermilliod:2006}, the $JHK_S$ magnitudes from \twomass\ \citep{2MASS}, the W1--W3 magnitudes from \wise\ \citep{WISE}, and the $G$, $G_{\rm BP}$, and $G_{\rm RP}$ magnitudes from \gaia. Together, the available photometry spans the full stellar SED over the wavelength range 0.2--12~$\mu$m (see Figure~\ref{fig:sed}).  

We performed a fit using Kurucz stellar atmosphere models \citep{Kurucz1993}, with the effective temperature ($T_{\rm eff}$), surface gravity ($\log g$) and metallicity ([Fe/H]) adopted from the spectroscopically determined values. The remaining free parameter is the extinction $A_V$, which we limited to the maximum line-of-sight value from the Galactic dust maps of \citet{Schlegel:1998}. The resulting fit (Figure~\ref{fig:sed}) has a reduced $\chi^2$ of 1.5, with best-fit $A_V = 0.20 \pm 0.05$. Integrating the (unreddened) model SED gives the bolometric flux at Earth, $F_{\rm bol} = 4.09 \pm 0.14 \times 10^{-10}$ erg~s$^{-1}$~cm$^{-2}$. Taking the $F_{\rm bol}$ and $T_{\rm eff}$ together with the \gaia\ parallax, gives the stellar radius, $R_\star = 0.846 \pm 0.023~R_\odot$. In addition, we can estimate the stellar mass from the empirical relations of \citet{Torres:2010}, giving $M_\star = 0.89 \pm 0.03~M_\odot$, consistent with the (less precise) value obtained from $R_\star$ together with the spectroscopic $\log g$, $0.74 \pm 0.17~M_\odot$. 

\subsection{Isochrones Fitting}\label{sec:isochrones}

To derive a self-consistent set of physical parameters for the host star, we fit the observed properties of TOI-4010 to MIST evolutionary models \citep{Dotter2016, Choi2016} using the stellar model grid package \texttt{isochrones} \citep{Morton2015}. We defined a single star model using parallax from \gaia\ DR3, observed magnitudes ($J$, $H$, $K_{s}$, $G$, $G_{BP}$, $G_{RP}$, $W1$, $W2$, and $W3$), and $T_{\rm eff}$ and [Fe/H] from the HARPS-N SPC fit. Following the convention of \citet{Eastman2017}, we inflated the magnitude uncertainties to 0.02 for the \gaia\ bands and 0.03 for the \wise\ bands.

Our fit parameters were stellar mass, age, metallicity, distance, and extinction. We used the default priors from \texttt{isochrones}, except for a broad flat prior for metallicity.\footnote{Our \texttt{isochrones} priors were as follows. Mass: power law prior with power-law index $\alpha = -2.35$, bounded between 0.1 and 10 $M_{\odot}$; Age: flat log prior between 9 and 10.15; Metallicity: uniform prior between -2 and 10 dex; Distance: power law prior with $\alpha = 2$, bounded between 0 and 3000 pc; Extinction: uniform prior between 0 and 1.} We ran the fit using the \texttt{emcee} ensemble sampler \citep{ForemanMackey2013,Goodman2010} with 100 walkers for 40,000 steps, and considered the fit converged based on the chain being more than 50 times longer than the estimated autocorrelation time. We discarded the first 500 steps as burn-in.

Our \texttt{isochrones} fit concluded that TOI-4010 is a metal-rich ($[$Fe/H$] = 0.37\pm0.07$ dex) K-dwarf ($T_{\rm eff} = 4960\pm36$ K, $\log g~\text{(cgs)} = 4.54\pm0.02$) with mass $M_{\star} = 0.88\pm0.03~M_{\odot}$, and radius $R_{\star} = 0.83\pm0.01~R_{\odot}$. All properties are within 1$\sigma$ of those derived from the HARPS-N SPC and SED analysis. We defer to the \texttt{isochrones}-fitted parameters when requiring stellar parameters in the remainder of this paper.

\subsection{Stellar Activity}\label{sec:activity}

The HARPS-N DRS pipeline computes chromospheric activity indicators from each spectrum, including the Mount-Wilson S-index \cite[$S_{mw}$;][]{Noyes1984,Vaughan1978}, H$\alpha$-Index \citep{Cincunegui2007,Bonfils2007}, and Na-Index \citep{Diaz2007}, which measure emission in the cores of the Ca II H and K, H$\alpha$, and Na I D1 and D2 lines, respectively. The DRS also provides indicators that quantify changes in the mean line profile, namely the full full width at half maximum (FWHM) and bisector inverse slope \citep[BIS;][]{Queloz2001} of the CCF. Prior to further analysis, we removed one negative (and therefore unphysical) measurement from the $S_{mw}$ time series.

\begin{figure*}
    \centering
    \includegraphics[width=\linewidth]{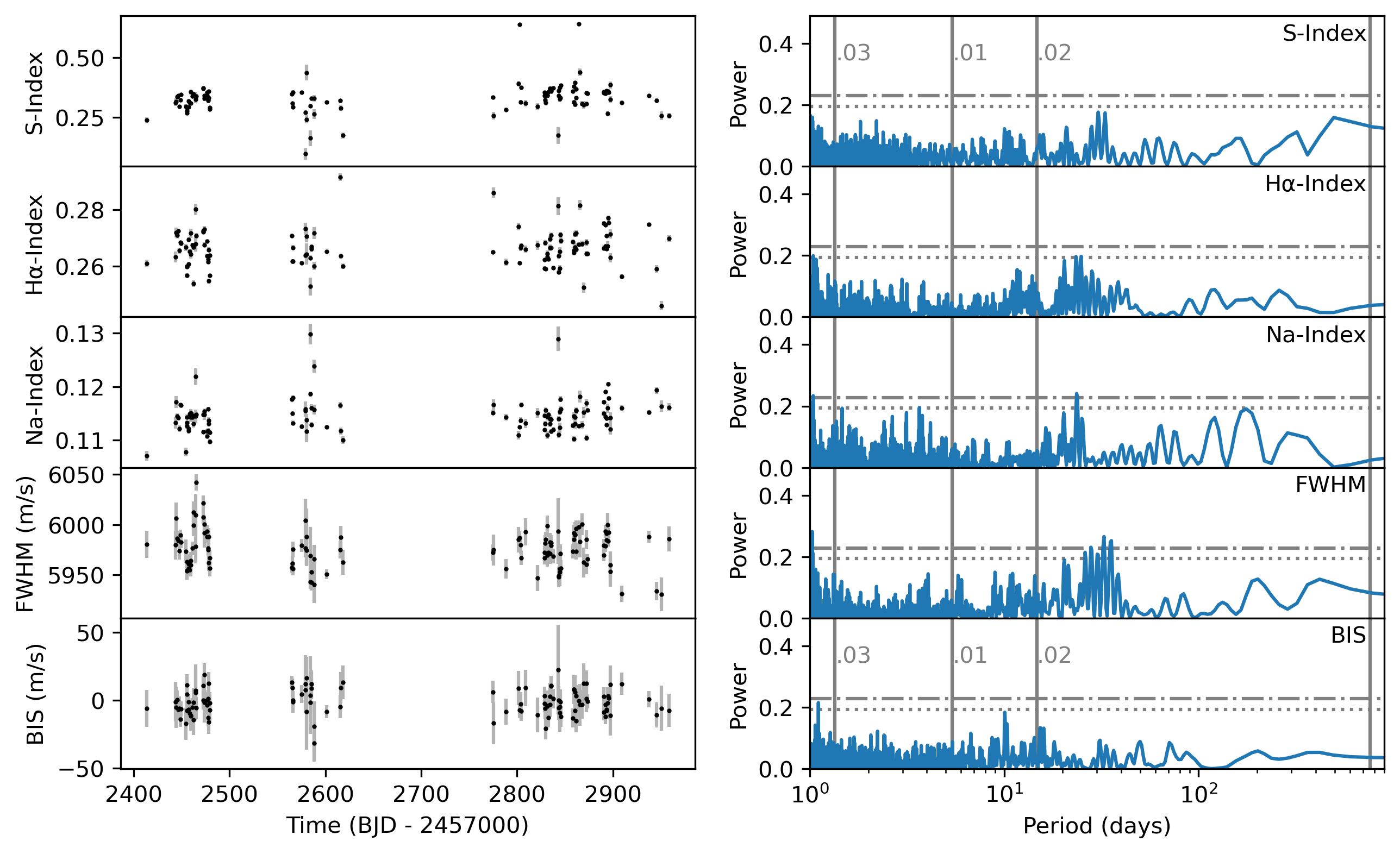}
    \caption{\textbf{Left:} Time series of activity indicators from HARPS-N spectra. \textbf{Right:} GLS periodograms of activity indicators. The horizontal lines mark 1\% (dotted line) and 0.1\% (dash-dotted line) false alarm probability levels, while vertical lines mark the orbital periods of the three planets. The line at $P = 762$ days marks the possible orbital period of the fourth planet that was uncovered in our later RV analysis (see \S\ref{sec:RV}).}
    \label{fig:gls}
\end{figure*}

We computed generalized lomb-scargle (GLS) periodograms \citep{Zechmeister2009} for all indicators (Figure \ref{fig:gls}). There are peaks just above 0.1\% false alarm probability (FAP) levels at $P \sim 24$ days and $P \sim 33$ days in Na-Index and FWHM, respectively. However, we find no significant signals at the periods of the transiting planets. There are also no clear periodicities shared between periodograms, and the time series of the indicators show no long-term trends over the 545-day baseline.

We also searched for correlations between indicators and the RVs, recognizing that signals from activity indicators may not necessarily be periodic. In particular, anticorrelation between BIS and RVs can be considered a signature of short-term photospheric variability \citep{Queloz2001}. We find a Pearson rank correlation coefficient of $r = -0.16$, indicating no significant correlation between BIS and RVs. All other indicators feature weak or non-existent correlations ($|r| < 0.3$) with the RVs.

Overall, TOI-4010 appears to have low levels of stellar activity, and we conclude that activity is unlikely to influence our measurements of the planet masses.

\subsubsection{Stellar Rotation}

\begin{figure*}[t]
    \centering
    \includegraphics[width=\linewidth]{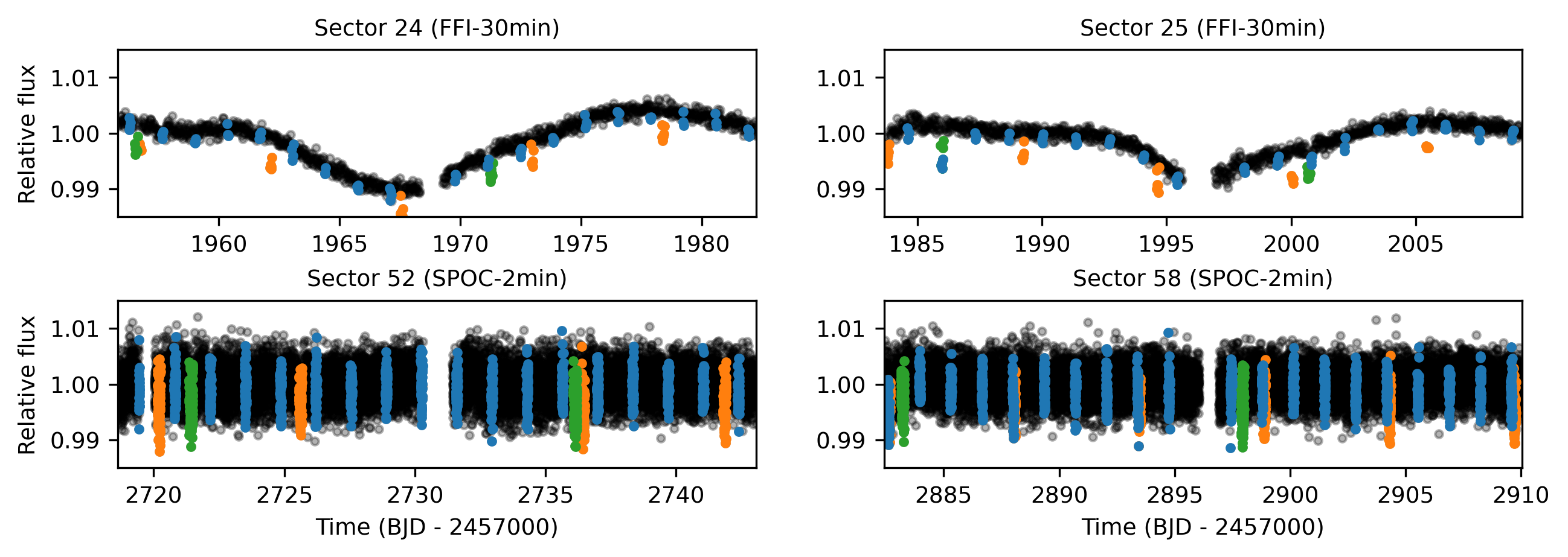}
    \caption{TESS light curve for TOI-4010 before removal of stellar variability. The transits of TOI-4010.01, .02, and .03 are indicated by blue, green, and orange dots, respectively, based on the transit ephemerides found in \S\ref{sec:joint}.}
    \label{fig:lightcurve}
\end{figure*}

A rotation signal with a period of $P_{\text{rot}} \sim 30$ days may be visible in the TESS photometry in Sectors 24 and 25 (Figure \ref{fig:lightcurve}). However, the star appears quieter in Sectors 52 and 58, and measuring rotation periods longer than $\sim13$ days using TESS is challenging due to the telescope orbit \citep[e.g.][]{Hedges2020,Claytor2022}.

Lacking strong periodicities in the activity indicators, we also cannot use the GLS periodograms to find the stellar rotation signal. Instead, we estimate $P_{\text{rot}}$ by relating it to $\log{R^{\prime}_{HK}}$, the logarithmic fraction of the flux in the H and K line cores to photometric contributions from the star. We use the S-index and $B-V$ colour of the star to estimate $\log{R^{\prime}_{HK}}$ using the method outlined by \citet{Noyes1984}. Based on the weighted mean of $\langle S_{mw}\rangle = 0.34\pm0.06$ from the HARPS-N observations and color of $B - V = 1.09\pm0.07$ mag from the Fourth US Naval Observatory CCD Astrograph Catalog \citep[UCAC4;][]{UCAC4}, we find $\log{R^{\prime}_{HK}} = -4.90\pm0.24$, consistent with a low-activity star. This $\log{R^{\prime}_{HK}}$ suggests a stellar rotation period of $P_{\text{rot}} = 37.7\pm15.0$ days based on the relationship for K stars provided by \citet{Mascareno2016}, which does not correspond to any of the planet orbital periods.

\section{Planet Parameters}\label{sec:planets}

\subsection{Transit-Only Fit}

We first fit our TESS Sector 24 and 25 FFI-extracted light curves, the Sector 52 and 58 SPOC light curves, and ground-based photometry with a three-planet transit model to get precise ephemerides for each planet. We included $ip$-band LCOGT observations with clean full transits (TOI-4010.01 with LCO-McD on 2021 August 28; TOI-4010.02 with LCO-Teide on 2021 August 30 and October 27; TOI-4010.03 with LCO-Teide on 2022 August 06, 2022 August 24, and 2022 September 06) and the $I_{c}$-band OAA observations with the full transit of TOI-4010.01.

The model was parameterized by the orbital period ($P$), transit epoch ($T_{0}$), planet-to-star radius ratio ($R_{p}/R_{\star}$), and impact parameter ($b$) of each planet, as well as host star mass and radius. We adopted a quadratic limb-darkening law parameterized by $q_{1}, q_{2}$ from \citet{exoplanet:kipping13}, with limb-darkening parameters fit separately for TESS, LCO, and OAA bands. We fit for flux offsets to the FFI and SPOC light curves separately as well as each night of ground-based photometry, and jitter terms added in quadrature to the errors of all observations. We assumed circular orbits, under the expectation that our subsequent fits including HARPS-N RV data would provide significantly better constraints on orbital eccentricity ($e$) and argument of pericenter ($\omega$) than transit photometry alone.

To evaluate the model's joint posterior, we used \texttt{PyMC3} \citep{exoplanet:pymc3} within the \texttt{exoplanet} package \citep{exoplanet:joss}. We sampled four chains for 4000 tuning steps and 4000 draw steps each, and confirmed that the chains converged according to the Gelman-Rubin convergence statistic for each parameter satisfying $\hat{r} < 1.01$ \citep{Gelman1992}. 

\begin{figure}[b]
    \centering
    \includegraphics[width=\linewidth]{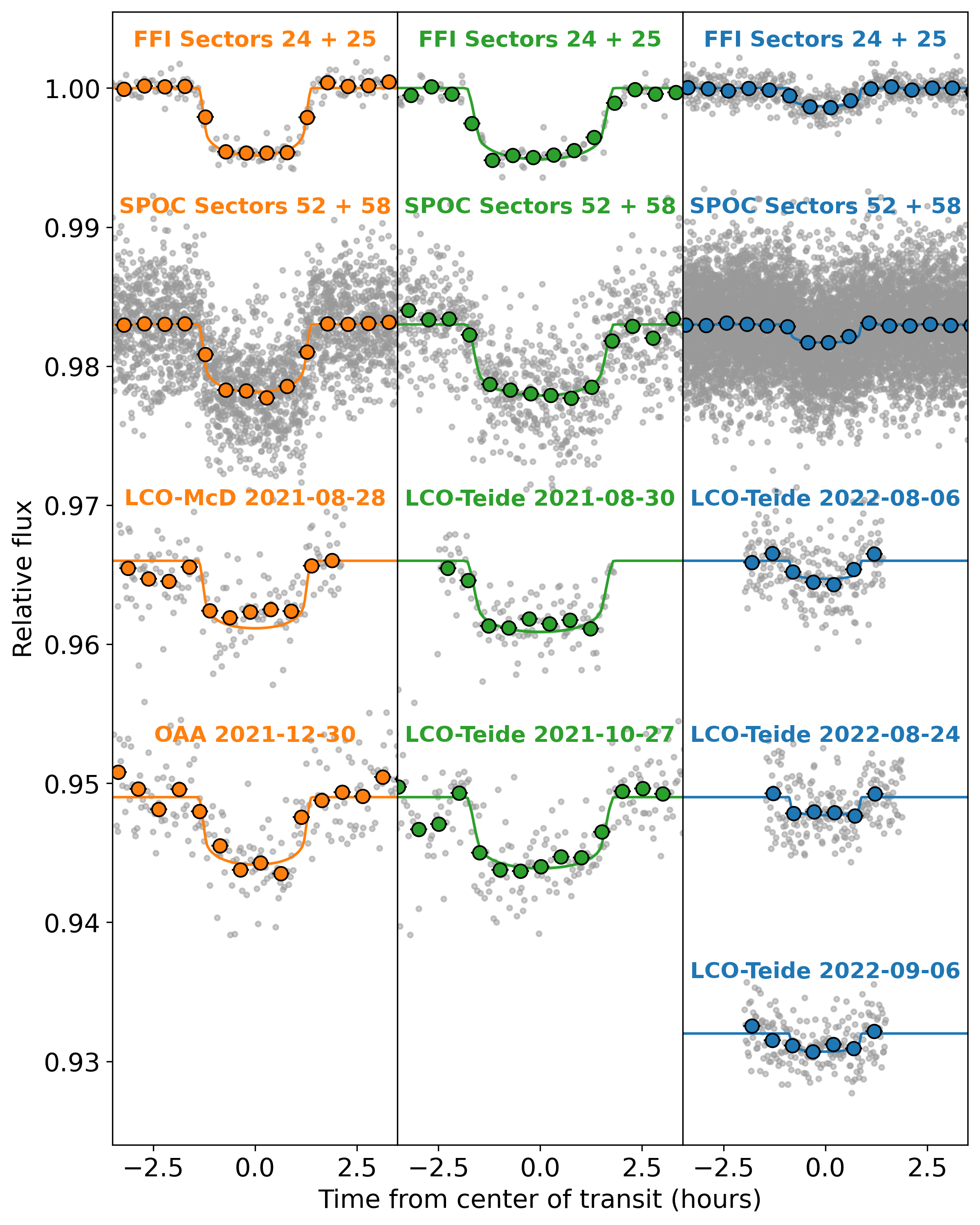}
    \caption{Photometric observations from TESS, LCO, and OAA, folded on the transit-only model fit periods of TOI-4010.01 (orange), 4010.02 (green), and 4010.03 (blue). The grey points reflect the photometric observations while coloured circles are data averaged in 30 min bins. The solid lines show the median transit model. The phase diagrams are offset by an arbitrary flux amount to improve visibility.}
    \label{fig:photometry}
\end{figure}

The transit phase diagrams are plotted in Figure \ref{fig:photometry}. Due to having observations spanning 954 days, the orbital periods and transit epochs are known precisely for all three planets. We improved the uncertainties in $P$ by two orders of magnitude with respect to the Sector 24 and 25 QLP values for all three candidates. The posterior distributions of $P$ and $T_{0}$ from this fit are indistinguishable from the results of the final joint model, so we show only the latter
results (see \S\ref{sec:joint}) for simplicity.

\subsection{RV-Only Fit}\label{sec:RV}

We separately fit the HARPS-N RV observations to ensure each planetary signal was present in the data. Visual inspection of the RV data revealed a significant long-term trend, resulting in a peak-to-peak change of $\sim$100 m~s$^{-1}$ across the observations (Figure \ref{fig:harpsn}). While this could be modeled by a simple quadratic trend in time, we do not neglect the possibility that it could be better modeled as a Keplerian due to a long-period giant planet companion. Furthermore, the stellar activity indicators do not show significant signals near the periods expected for the long-period planet (\S\ref{sec:activity}), so we do not believe the signal is activity-related. Therefore, we consider five classes of models:

\begin{figure}[t!]
    \centering
    \includegraphics[width=\linewidth]{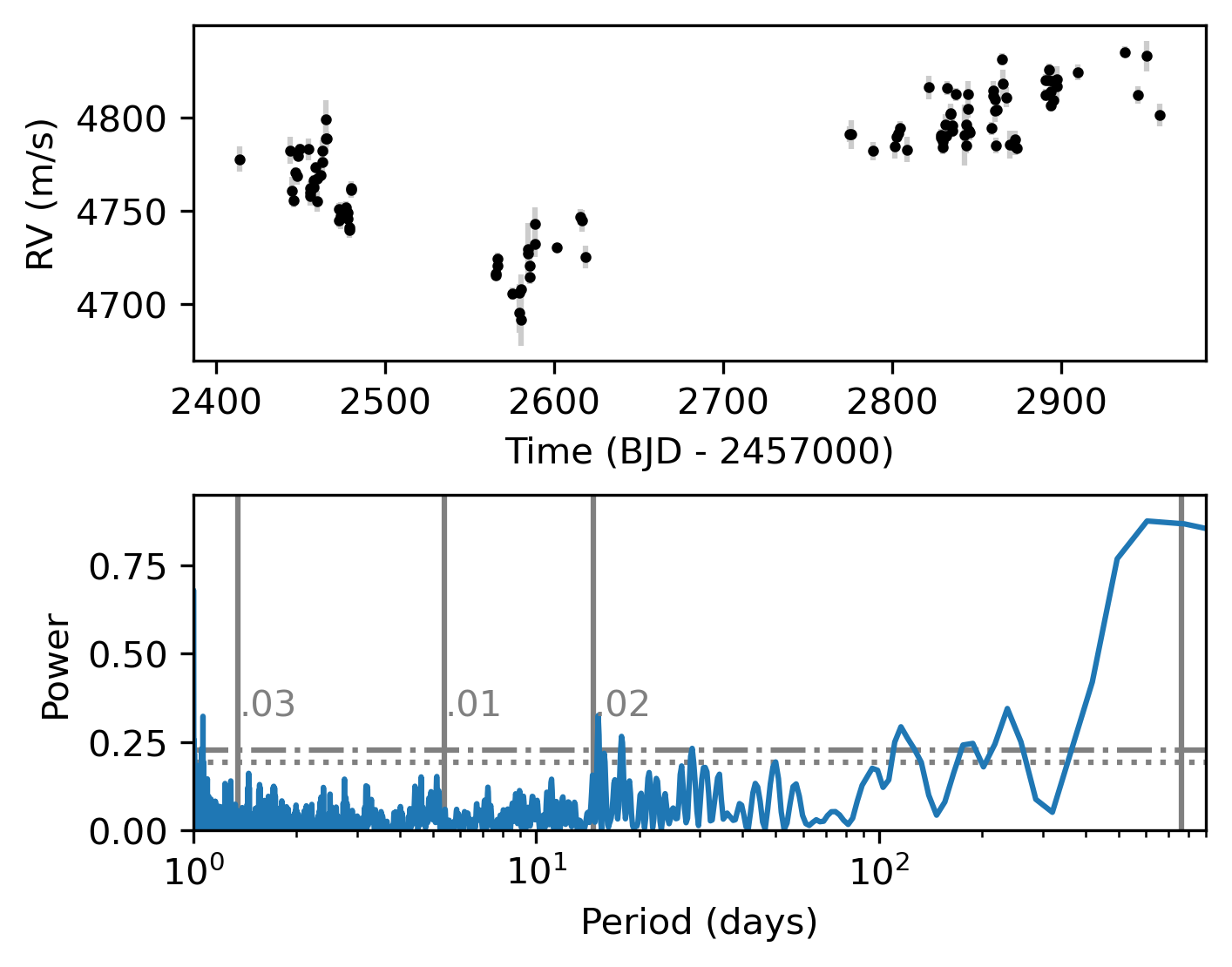}
    \caption{\textbf{Top:} 112 HARPS-N RV observations of TOI-4010. \textbf{Bottom:} The GLS periodogram of the RVs using the same format as Figure \ref{fig:gls}. There is a long-period signal with a semi-amplitude of $\sim50$ m s$^{-1}$ that dominates, indicating a fourth planet.}
    \label{fig:harpsn}
\end{figure}

\begin{itemize}
    \item 0p (``no-planet'') model: a quadratic trend,
    \item 1p (``one-planet'') models: a quadratic trend + a single Keplerian,
    \item 2p (``two-planet'') models: a quadratic trend + two Keplerians,
    \item 3p (``three-planet'') model: a quadratic trend + three Keplerians, and
    \item 4p (``four-planet'') model: four Keplerians.
\end{itemize}
 
For one- and two-planet models, we tried all possible combinations of the three transiting planets. For the four-planet models, we set wide uniform priors on the period and mid-transit time of the fourth planet. To test the sensitivity of our fit to choice of prior, we let the period vary between $500 - 600$ days, $600 - 700$ days, $700 - 800$ days, $800 - 900$ days, and $500 - 900$ days in five separate fits. While the correct orbital period may be longer than 900 days, fits with priors covering longer periods often failed to converge, likely due to the complexity of the system and the lack of sufficient observations to fit an eccentric orbit with a period much longer than the timespan of the data (545 days).

We fixed the periods and transit times of each transiting planet equal to the median of the transit-only fit results, under the expectation that the transit photometry should provide orders of magnitude more precise measurements of these properties than the RV data alone could constrain. This decision significantly sped up the computational time of the model fits. We also relaxed the circular orbit assumption by fitting for the eccentricity and argument of pericenter of each planet's orbit. For eccentricity priors, we used the distribution from \citet{VanEylen2019} for the three inner planets, and \citet{Kipping2013a} for the long-period giant planet. The argument of pericenter was constrained to the range $-\pi$ to $\pi$, and the sampling was performed in two-dimensional vector space ($\sqrt{e}\sin{\omega}$, $\sqrt{e}\cos{\omega}$) to avoid the sampler seeing a discontinuity at values of $\pi$. All models fit for an RV offset and a jitter term which was added in quadrature to the uncertainties. Because the activity indicators for TOI-4010 showed low levels of stellar activity and no strong periodicities, we opted not to fit the RV observations with a more complicated noise model.

We compared the models based on the leave-one-out cross-validation information criterion \cite[LOO;][]{Vehtari2015} as computed from the \texttt{exoplanet} results using the \texttt{arviz} package for analysis of Bayesian models \citep{arviz_2019}. Unlike simpler estimates for predictive error such as the Bayesian information criterion (BIC) or Akaike information criterion (AIC), which are based on point estimates of the maximum likelihood for the model, LOO is a fully Bayesian model selection method that uses the entire posterior distribution. A model is favoured if it has a higher log(LOO) score. As shown in Table \ref{tab:RVonly}, the three-planet model is a better fit to the data compared to all zero-, one-, and two-planet models, with $\Delta \log{(\text{LOO})} = 2.7\pm2.6$ over the next best model (the two-planet model featuring TOI-4010.01 and .02). Combined with the fact that ground-based photometry detected all three transiting signals on-target, high-resolution imaging follow-up did not find evidence for an unresolved eclipsing binary star, and none of the periods of the planets are co-incident with high levels of stellar activity, we consider that the transiting planet candidates from TESS are bonafide planets as confirmed by independent radial velocity measurements. We refer to TOI-4010.01, .02, and .03 as TOI-4010 c, d, and b, respectively, ordered alphabetically by increasing orbital period.

\begin{table*}[t]
    \centering
    \caption{Results from fitting the HARPS-N RV data alone, testing no-planet (0p), one-planet (1p), two-planet (2p), three-planet (3p), and four-planet (4p) models. We set the periods of the transiting planets fixed to the transit-only fit results, and used wide uniform priors for the period of the fourth planet in the 4p model. A model is favoured if it has a higher log(LOO) score, demonstrating that the 4p models are the best match to the data. The root-mean-square (RMS) error  for each scenario was computed from the difference between the data and the median RV model fit.}
    \begin{tabular}{ccccccc}
    \hline\hline
        Model & Planet & Period Prior & $K$ (m~s$^{-1}$) & $e$ & log(LOO) & RMS Error (m~s$^{-1}$) \\
    \hline
        0p & - & - & - & - & $-479.7\pm9.5$ & $17.2$ \\
        \hline
        1p & TOI-4010.01 & $\mathcal{F}(5.4147)$\footnote{$\mathcal{F}(x)$ denotes a value fixed at $x$} & $7.8_{-2.2}^{+2.2}$ & $0.04_{-0.03}^{+0.04}$ & $-474.7\pm10.1$ & $16.3$\\
        \hline
        1p & TOI-4010.02 & $\mathcal{F}(14.7089)$ & $8.6_{-2.4}^{+2.5}$ & $0.04_{-0.03}^{+0.05}$ & $-474.1\pm8.3$ & $16.1$ \\
        \hline
        1p & TOI-4010.03 & $\mathcal{F}(1.3483)$ & $4.3_{-2.1}^{+2.1}$ & $0.03_{-0.02}^{+0.04}$ & $-478.8\pm9.2$ & $16.9$\\
        \hline
        2p & TOI-4010.01 & $\mathcal{F}(5.4147)$ & $7.6_{-2.1}^{+2.1}$ & $0.03_{-0.02}^{+0.04}$ & $-468.6\pm9.0$ & $15.2$\\
        & TOI-4010.02 & $\mathcal{F}(14.7089)$ & $8.5_{-2.2}^{+2.3}$ & $0.04_{-0.03}^{+0.06}$ \\
        \hline
        2p & TOI-4010.01 & $\mathcal{F}(5.4147)$ & $7.9_{-2.1}^{+2.2}$ & $0.04_{-0.02}^{+0.04}$ & $-473.5\pm10.0$ & $15.9$\\
        & TOI-4010.03 & $\mathcal{F}(1.3483)$ & $4.4_{-2.0}^{+2.1}$ & $0.04_{-0.03}^{+0.04}$\\
        \hline
        2p & TOI-4010.02 & $\mathcal{F}(14.7089)$ & $9.1_{-2.3}^{+2.4}$ & $0.04_{-0.03}^{+0.06}$ & $-472.2\pm8.1$ & $15.7$\\
        & TOI-4010.03 & $\mathcal{F}(1.3483)$ & $5.1_{-2.1}^{+2.1}$ & $0.03_{-0.02}^{+0.04}$ \\
        \hline
        3p & TOI-4010.01 & $\mathcal{F}(5.4147)$ & $7.7_{-2.0}^{+2.0}$ & $0.03_{-0.02}^{+0.04}$ & $-465.9\pm8.9$ & $14.8$\\
        & TOI-4010.02 & $\mathcal{F}(14.7089)$ & $8.9_{-2.1}^{+2.3}$ & $0.04_{-0.03}^{+0.05}$\\
        & TOI-4010.03 & $\mathcal{F}(1.3483)$ & $5.1_{-1.9}^{+1.9}$ & $0.03_{-0.02}^{+0.04}$ \\
        \hline 
        4p & TOI-4010.01 & $\mathcal{F}(5.4147)$ & $8.0_{-0.8}^{+0.9}$ & $0.03_{-0.02}^{+0.03}$ & $-371.3\pm9.2$ & $6.5$ \\
        & TOI-4010.02 & $\mathcal{F}(14.7089)$ & $11.0_{-0.9}^{+0.9}$ & $0.07_{-0.03}^{+0.03}$ \\
        & TOI-4010.03 & $\mathcal{F}(1.3483)$ & $7.0_{-0.8}^{+0.8}$ & $0.03_{-0.02}^{+0.03}$\\
        & Planet 4 & $\mathcal{U}(500, 600)$\footnote{$\mathcal{U}(a,b)$ denotes a uniform prior between $a$ and $b$} & $49.1_{-1.2}^{+1.2}$ & $0.21_{-0.03}^{+0.03}$\\
        \hline
        4p & TOI-4010.01 & $\mathcal{F}(5.4147)$ & $8.0_{-0.8}^{+0.8}$ & $0.03_{-0.02}^{+0.03}$ & $-369.0\pm9.1$ & $6.4$\\
        & TOI-4010.02 & $\mathcal{F}(14.7089)$ & $10.9_{-0.9}^{+0.9}$ & $0.07_{-0.03}^{+0.03}$ \\
        & TOI-4010.03 & $\mathcal{F}(1.3483)$ & $6.9_{-0.8}^{+0.8}$ & $0.03_{-0.02}^{+0.03}$ \\
        & Planet 4 & $\mathcal{U}(600, 700)$ & $51.2_{-1.6}^{+1.7}$ & $0.22_{-0.03}^{+0.03}$ \\
        \hline
        4p & TOI-4010.01 & $\mathcal{F}(5.4147)$ & $8.0_{-0.8}^{+0.8}$ & $0.03_{-0.02}^{+0.03}$ & $-368.4\pm9.0$ & $6.3$\\
        & TOI-4010.02 & $\mathcal{F}(14.7089)$ & $10.8_{-0.9}^{+0.9}$ & $0.07_{-0.03}^{+0.03}$ \\
        & TOI-4010.03 & $\mathcal{F}(1.3483)$ & $6.9_{-0.8}^{+0.8}$ & $0.03_{-0.02}^{+0.03}$ \\
        & Planet 4 & $\mathcal{U}(700, 800)$ & $54.3_{-2.0}^{+2.1}$ & $0.26_{-0.03}^{+0.03}$ \\
        \hline
        4p & TOI-4010.01 & $\mathcal{F}(5.4147)$ & $8.0_{-0.8}^{+0.8}$ & $0.03_{-0.02}^{+0.03}$ & $-368.3\pm9.0$ & $6.3$\\
        & TOI-4010.02 & $\mathcal{F}(14.7089)$ & $10.8_{-0.9}^{+0.9}$ & $0.07_{-0.03}^{+0.03}$ \\
        & TOI-4010.03 & $\mathcal{F}(1.3483)$ & $6.9_{-0.8}^{+0.8}$ & $0.03_{-0.02}^{+0.03}$ \\
        & Planet 4 & $\mathcal{U}(800, 900)$ & $57.7_{-2.3}^{+2.5}$ & $0.29_{-0.03}^{+0.03}$ \\
        \hline
        4p & TOI-4010.01 & $\mathcal{F}(5.4147)$ & $8.0_{-0.8}^{+0.8}$ & $0.03_{-0.02}^{+0.03}$ & $-368.7\pm9.0$ & $6.3$\\
        & TOI-4010.02 & $\mathcal{F}(14.7089)$ & $10.9_{-0.9}^{+0.9}$ & $0.07_{-0.03}^{+0.03}$ \\
        & TOI-4010.03 & $\mathcal{F}(1.3483)$ & $6.9_{-0.8}^{+0.8}$ & $0.03_{-0.02}^{+0.03}$ \\
        & Planet 4 & $\mathcal{U}(500, 900)$ & $54.5_{-3.3}^{+3.5}$ & $0.26_{-0.04}^{+0.04}$ \\
    \end{tabular}
    \label{tab:RVonly}
\end{table*}

However, the four-planet models are significantly favoured over all other models, as indicated by differences in log(LOO) that are several times larger than the standard errors of the differences. The $500 - 900$-day prior gives $\Delta \log{(\text{LOO})} = 97.4\pm10.7$ over the three-planet model. Subtracting the median RV model from the observations also results in the lowest spread in residuals, with root-mean-square (RMS) error of $6.3$~m s$^{-1}$ compared to $14.8$~m s$^{-1}$ for the three-planet case. While the RV semi-amplitude and eccentricity of the fourth planet change slightly depending on the choice of period prior (from $K = 49.1_{-1.2}^{+1.2}$ m~s$^{-1}$ and $e = 0.21_{-0.03}^{+0.03}$ assuming a period between $500 - 600$ days, to $K = 57.7_{-2.3}^{+2.5}$ m~s$^{-1}$ and $e = 0.29_{-0.03}^{+0.03}$ assuming a period between $800 - 900$ days), all solutions point to a long-period giant planet companion in a moderately eccentric orbit. The fitted properties of the transiting planets also remain consistent across all four-planet fits, indicating that their solutions are robust. 

Using the widest prior ($500 - 900$ days) to best capture the uncertainties, we find that the fourth planet can be possibly described by $P = 760_{-90}^{+91}$ days, $K = 54.5_{-3.3}^{+3.5}$ m~s$^{-1}$, $e = 0.26_{-0.04}^{+0.04}$, and $M_{p}\sin{i} = 705_{-37}^{+41}~M_{\oplus}$. We consider this a new long-period planet, and refer to it as TOI-4010 e.

\begin{figure}[ht]
    \centering
    \includegraphics[width=\linewidth]{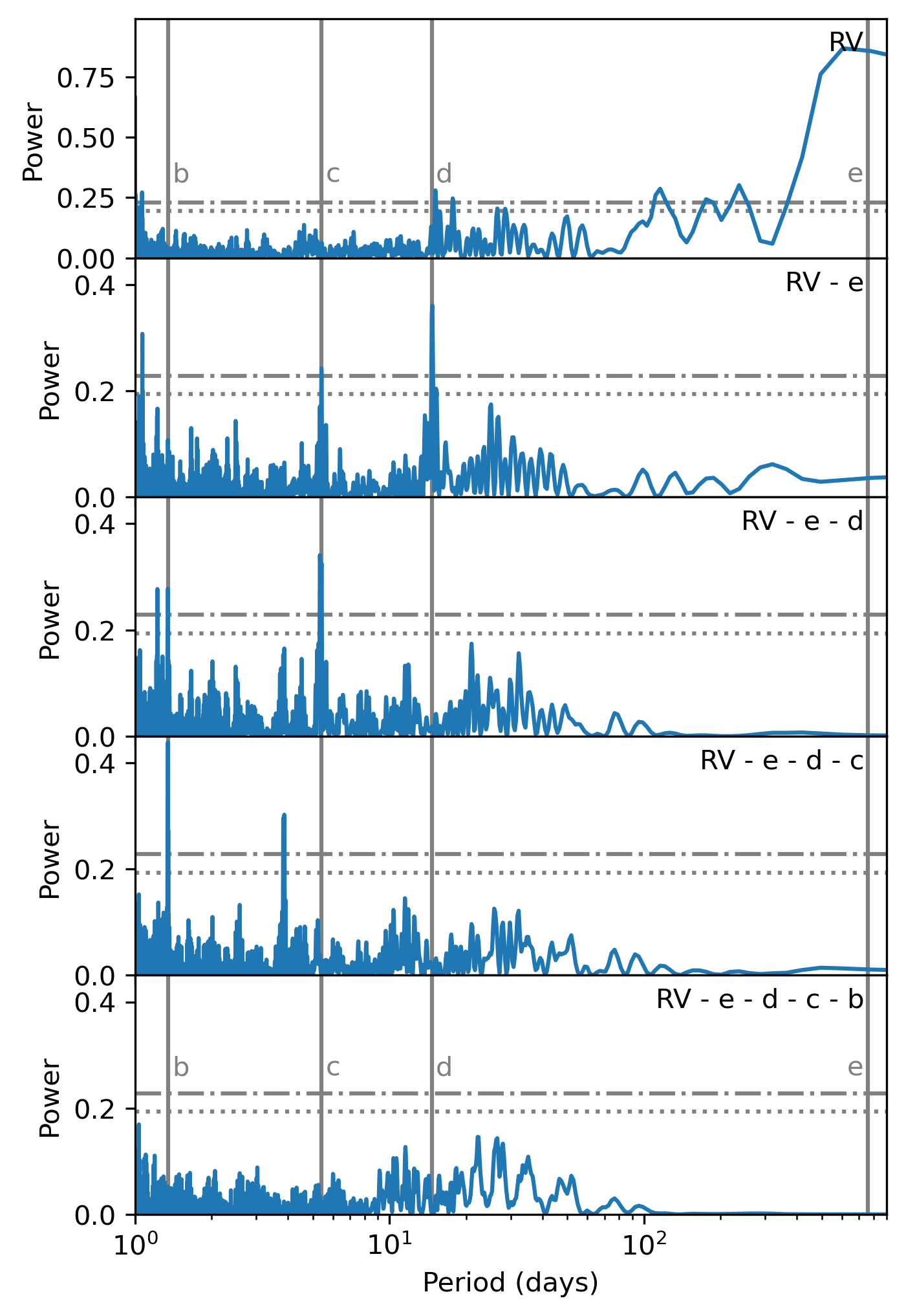}
    \caption{GLS peridograms of the HARPS-N RVs. The top panel shows periodograms of the original RVs, while subsequent panels show periodograms of the RVs after subtracting our median model fits for TOI-4010 e, d, c, and b, respectively. In each panel, the horizontal lines mark 1\% (dotted line) and 0.1\% (dash-dotted line) false alarm probability levels, while vertical lines mark the orbital periods of the four planets.}
    \label{fig:rvLS}
\end{figure}

Finally, we inspected the residuals with a GLS periodogram after subtracting the median RV model of each planet from the RVs. The residuals do not show any statistically significant periodicity that could be attributed to additional planets or stellar activity (Figure \ref{fig:rvLS}).

\subsection{Joint transit + RV Model Fit}\label{sec:joint}

To obtain the final parameters, we performed a joint analysis of the TESS photometry and HARPS-N RV data using \texttt{exoplanet}. Due to the significantly increased computational expense of the joint fit, we did not include ground-based photometry.

\begin{figure*}[ht]
    \centering
    \includegraphics[width=0.40\linewidth]{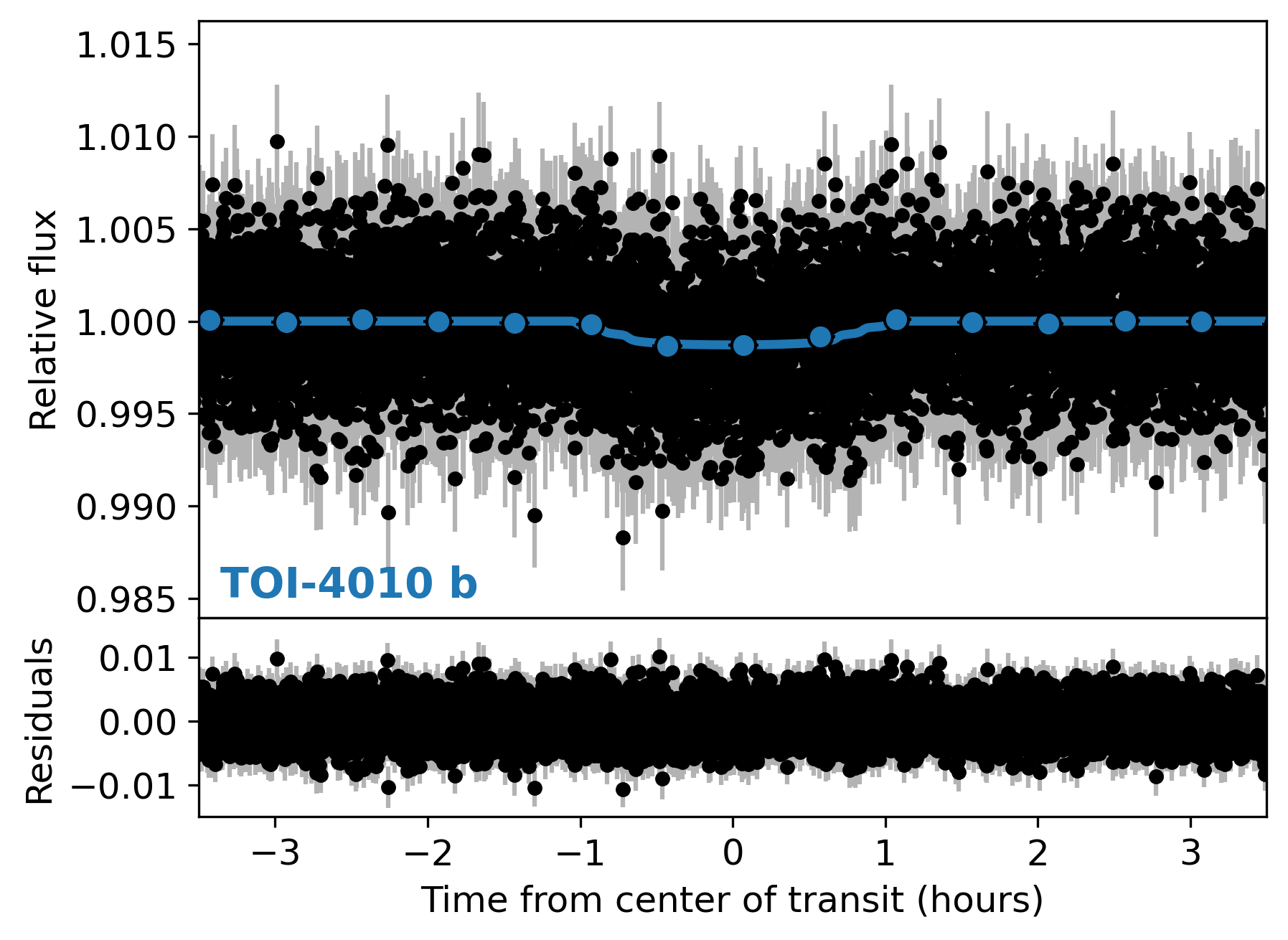}
    \includegraphics[width=0.40\linewidth]{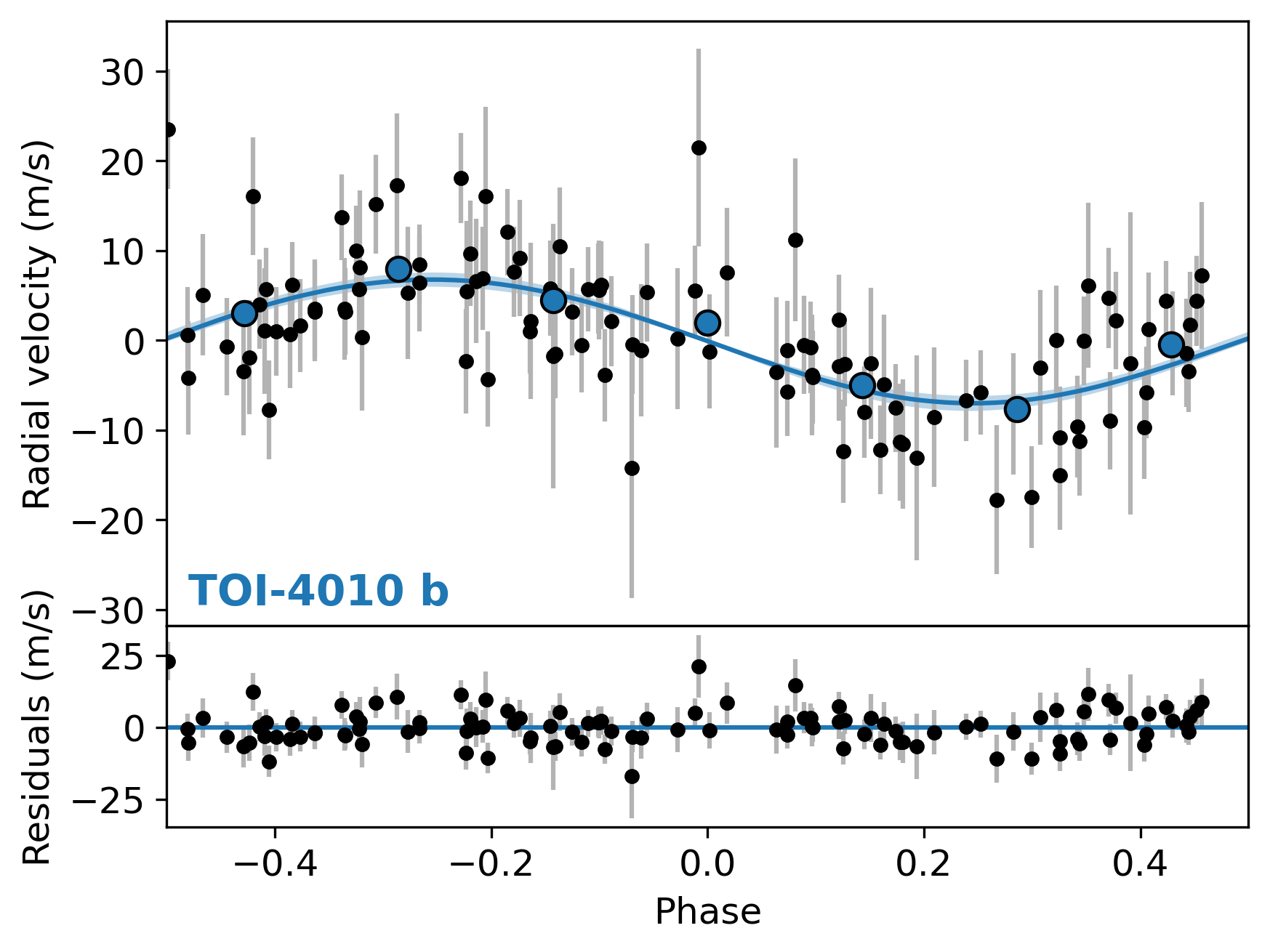}
    \includegraphics[width=0.40\linewidth]{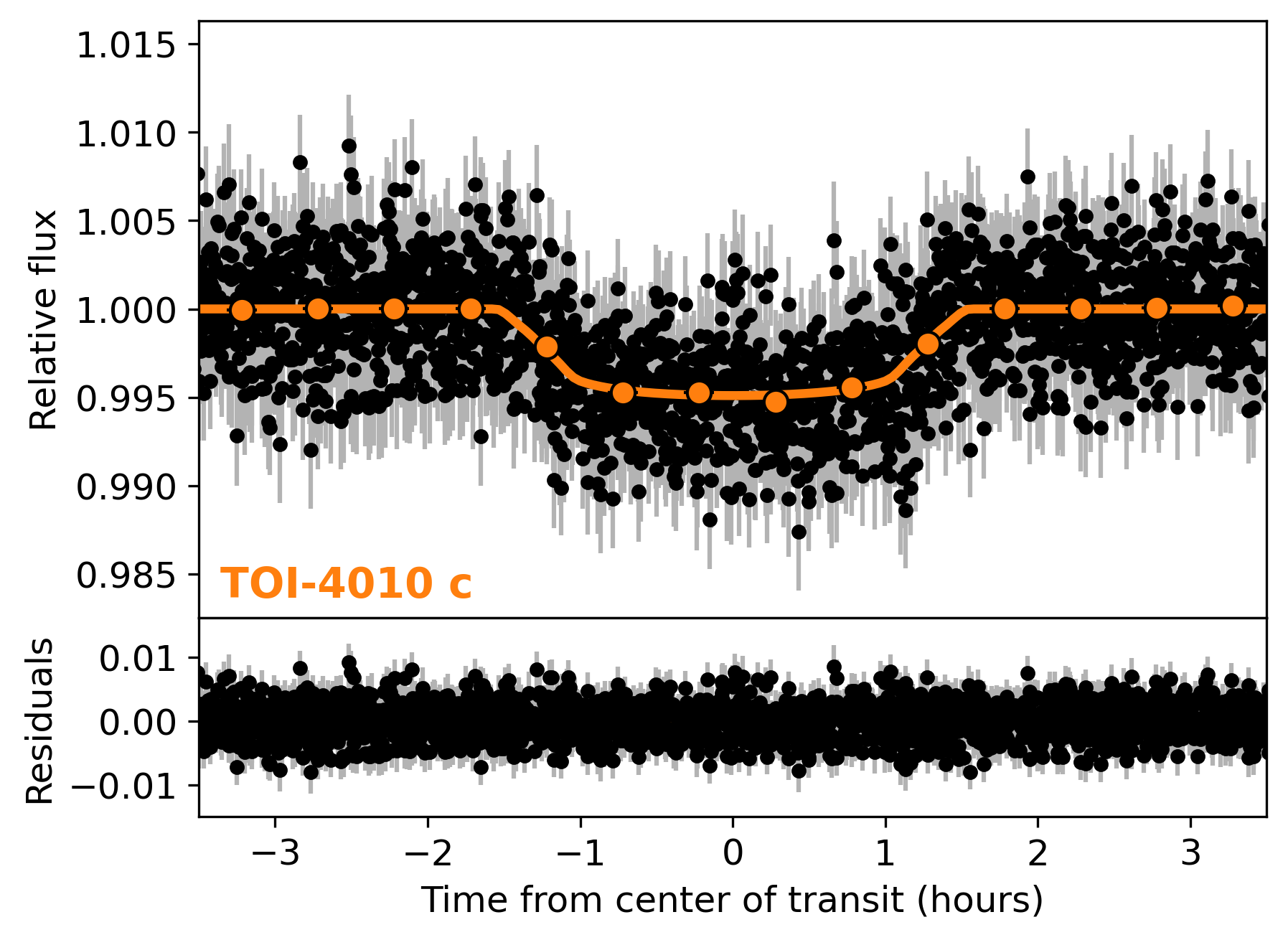}
    \includegraphics[width=0.40\linewidth]{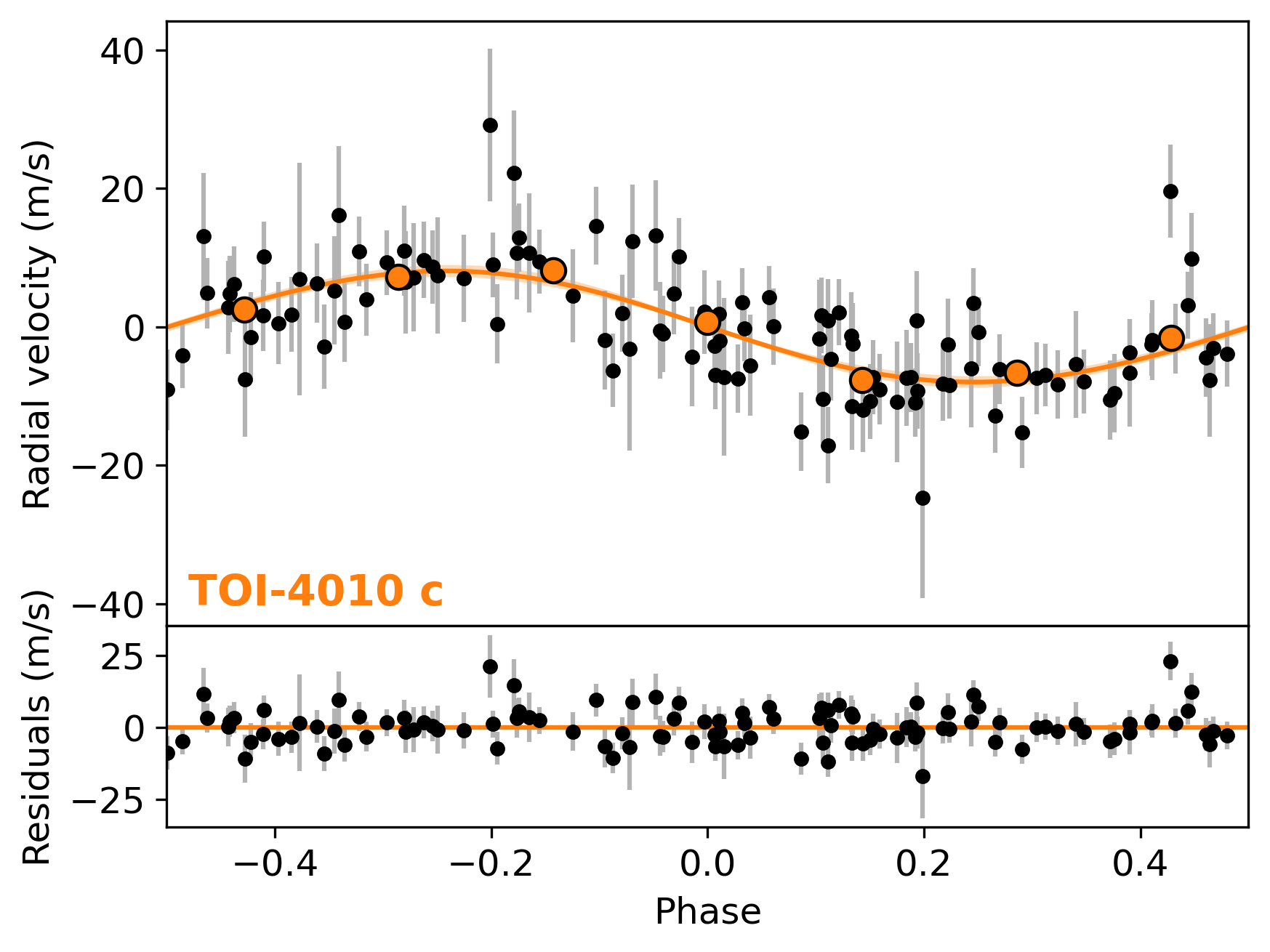}
    \includegraphics[width=0.40\linewidth]{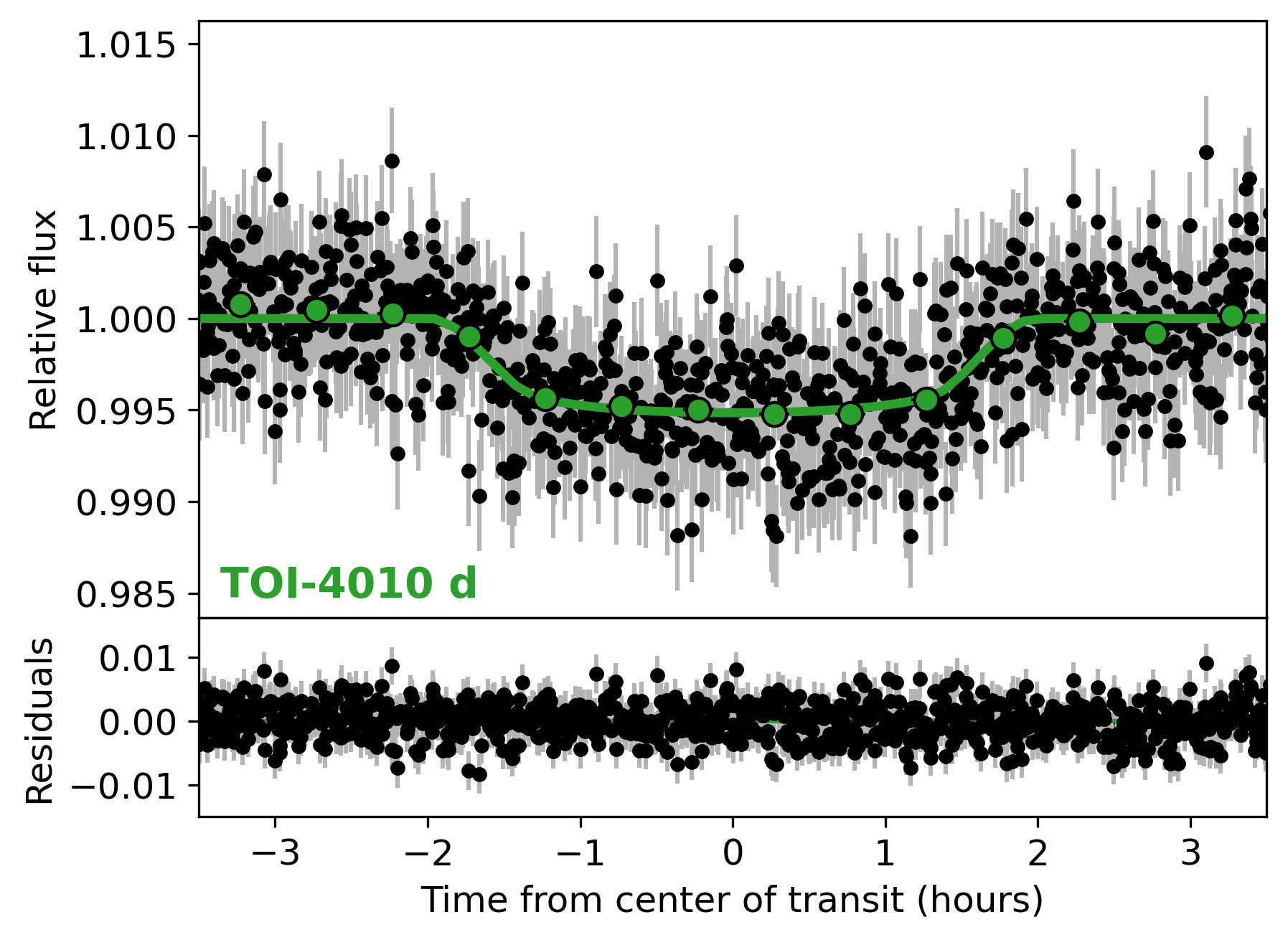}
    \includegraphics[width=0.40\linewidth]{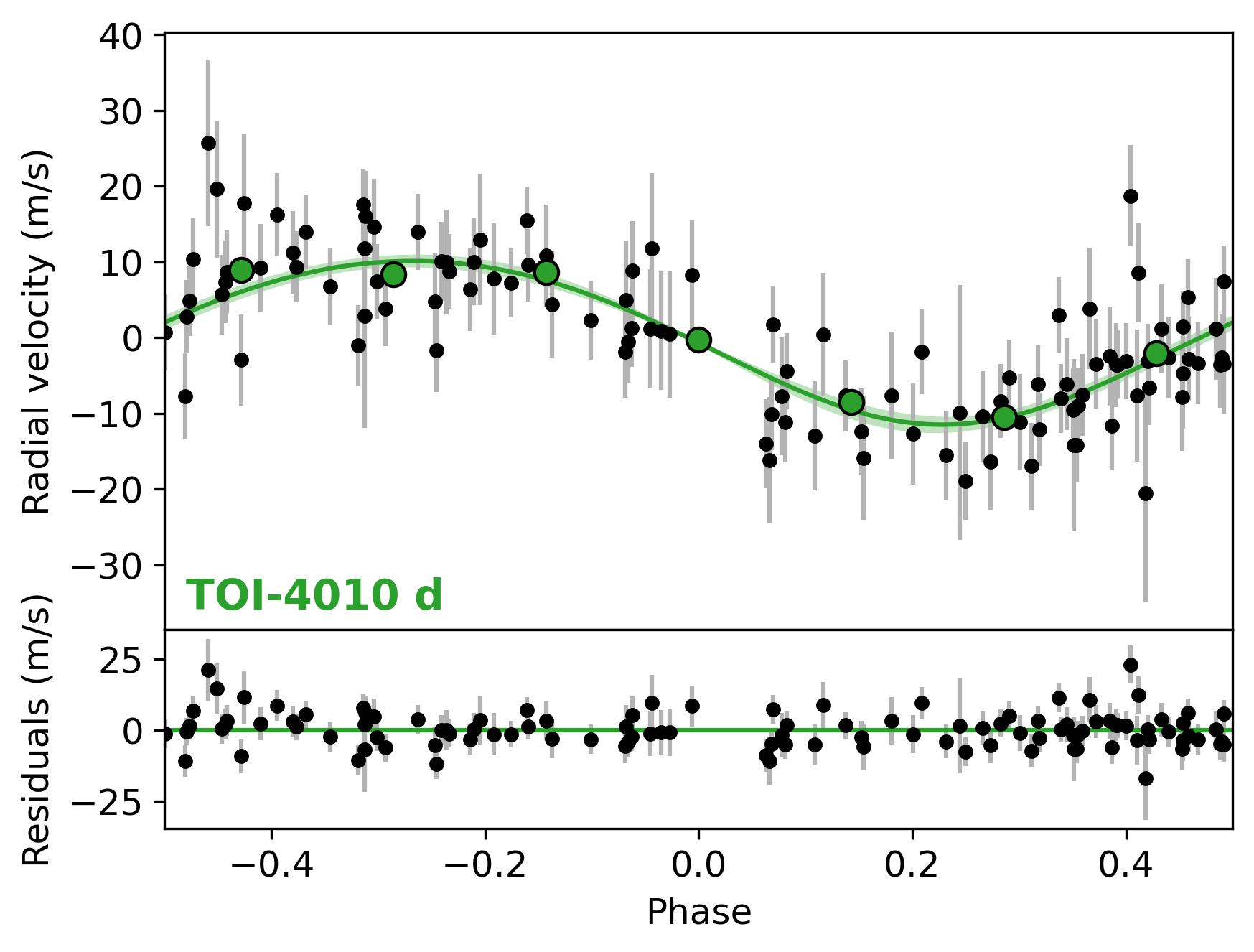}
    \caption{Plots of the final joint transit + RV model fits, with residuals after subtracting the median models provided in the lower panel of each phase diagram. Black points show observations with offsets subtracted and jitter terms added in quadrature with uncertainties, while coloured circles represent binned data. Solid coloured lines represent median model values, while shaded regions cover 16th to 84th percentiles. The transit (left) and RV (right) phase diagrams for each planet are shown in each row. In each planet's phase diagram, the best-fit models for the other three planets have been subtracted from the data.}
    \label{fig:finalfits}
\end{figure*}

\begin{figure}[ht]
    \centering
    \includegraphics[width=\linewidth]{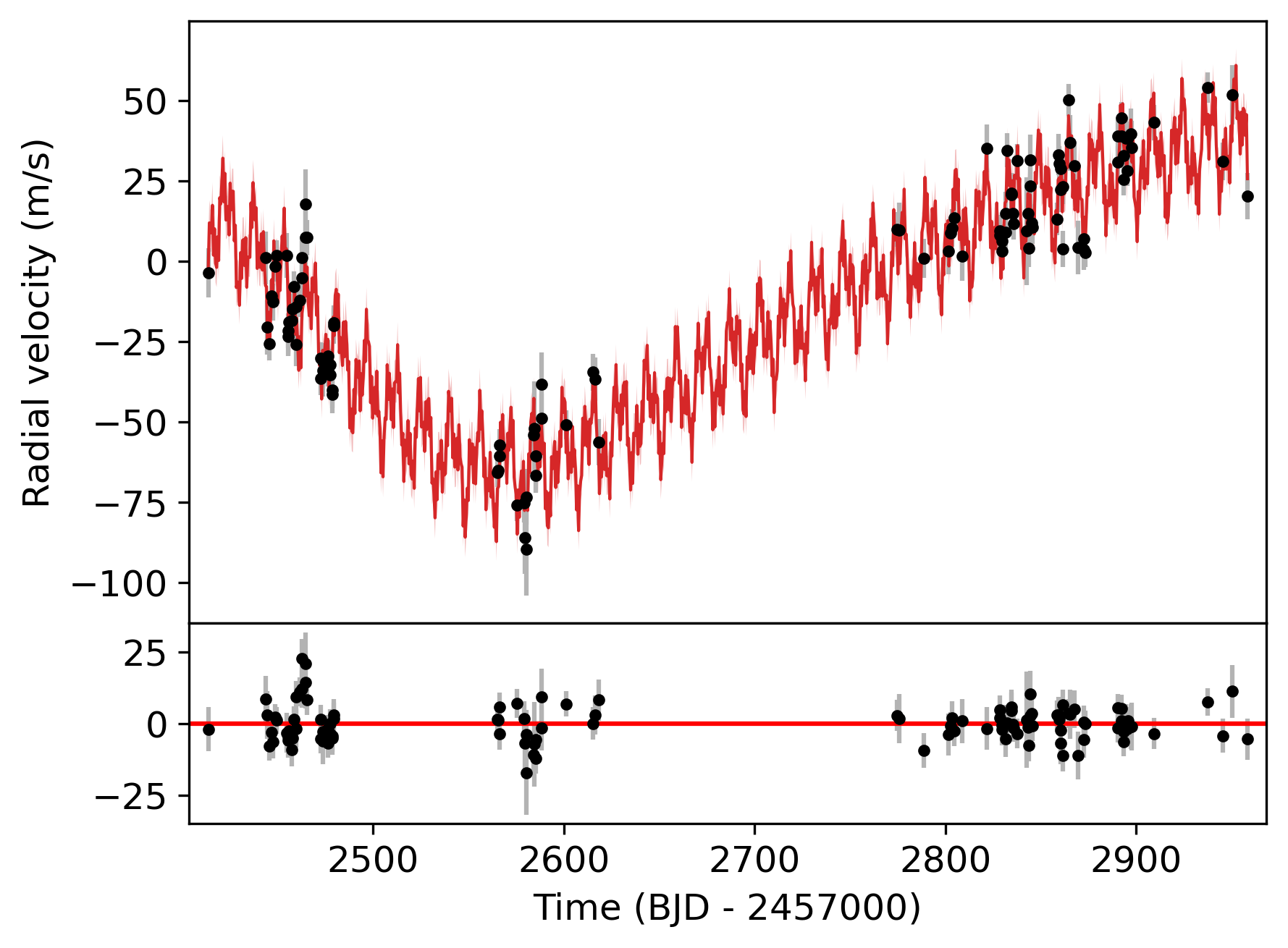}
    \includegraphics[width=\linewidth]{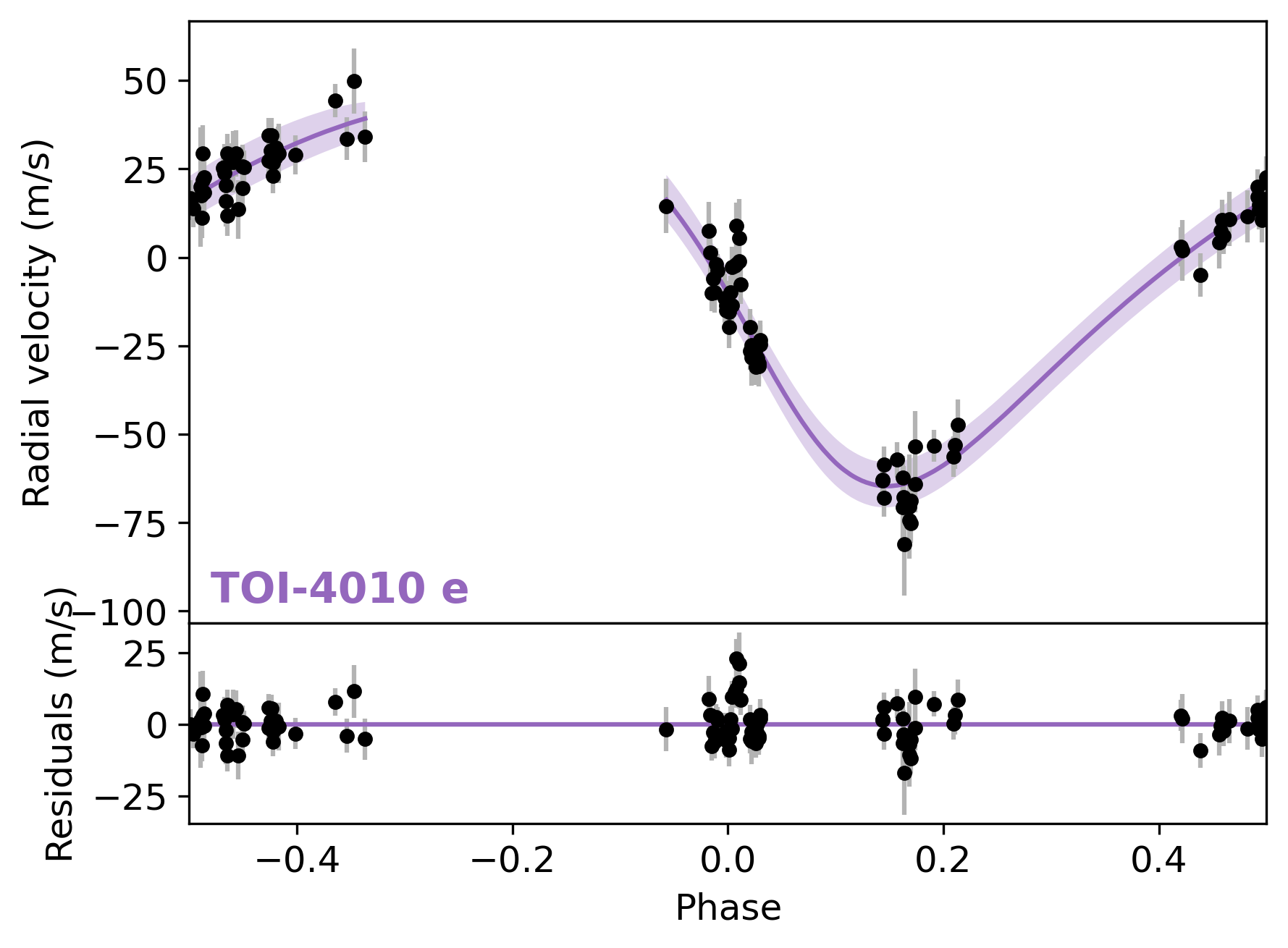}
    \caption{\textbf{Top:} HARPS-N RV time series, which has a long-term trend best explained by an outer giant planet. The four-planet model is over-plotted in red, and the residuals after subtracting the model are in the lower panel. \textbf{Bottom:} RV phase diagram for the long-period giant planet TOI-4010 e with the signals of the other three planets removed, in the same format as in Figure \ref{fig:finalfits}. The Keplerian model is plotted in purple. The gap in the phase diagram reflects the fact that the RV observing baseline is shorter than the estimated orbital period of the planet.}
    \label{fig:finalfits_e}
\end{figure}

We fit a four-planet model to the RVs based on the results of our RV-only fit, but only a three-planet model to the light curve due to the lack of transit for TOI-4010 e. Our adopted priors are in Table \ref{tab:priors}, and histograms of the distributions of all fit parameters after sampling four chains with 4000 tuning steps and 8000 draw steps are shown in Figure \ref{fig:posteriors}. The final fitted values are reported in Table \ref{tab:fits}, along with additional fundamental transit and physical properties derived from the fits. Because insolation flux and equilibrium temperature require knowledge of the host star's effective temperature, we drew random values from $T_{\rm eff}$ posterior from the \texttt{isochrones} results for each realization of the joint fit. Figure \ref{fig:finalfits} shows the TESS photometry and HARPS-N RV observations folded on the orbits of the three transiting planets, while Figure \ref{fig:finalfits_e} shows the full RV observations and phase diagram for the non-transiting planet.

\begin{table*}[h!]
    \centering
    \begin{tabular}{l|cccc|l}
    \hline\hline
    Parameter & TOI-4010 b & TOI-4010 c & TOI-4010 d & TOI-4010 e & Description \\
\hline\hline
\multicolumn{6}{c}{\textit{Planet Parameters}} \\
\hline
$P$ (days) & $1.348335_{-0.000002}^{+0.000003}$ & $5.414654_{-0.000007}^{+0.000007}$ & $14.70886_{-0.00003}^{+0.00003}$ & $762_{-90}^{+90}$ & Orbital period\footnote{Lower period limit for TOI-4010 e} (days) \\
$T_0$ (BTJD) & $2007.549_{-0.001}^{+0.001}$ & $2000.0523_{-0.0008}^{+0.0008}$ & $1985.985_{-0.001}^{+0.001}$ & $3218_{-86}^{+86}$ & Transit epoch (BJD - 2457000) \\
$K$ (m s$^{-1}$) & $6.9_{-0.8}^{+0.8}$ & $8.0_{-0.8}^{+0.8}$ & $10.9_{-0.9}^{+0.9}$ & $54.7_{-3.4}^{+3.6}$ & RV semi-amplitude (m~s$^{-1}$) \\
$R_p/R_\star$ & $0.033_{-0.001}^{+0.001}$ & $0.065_{-0.001}^{+0.001}$ & $0.068_{-0.002}^{+0.001}$ & - & Planet-to-star radius ratio \\
$b$ & $0.19_{-0.13}^{+0.12}$ & $0.31_{-0.15}^{+0.09}$ & $0.51_{-0.08}^{+0.06}$ & - & Impact parameter \\
$e$ & $0.03_{-0.02}^{+0.03}$ & $0.03_{-0.02}^{+0.03}$ & $0.07_{-0.03}^{+0.03}$ & $0.26_{-0.04}^{+0.04}$ & Orbital eccentricity \\
$\omega$ & $0.0_{-2.6}^{+2.7}$ & $1.0_{-2.1}^{+1.1}$ & $1.9_{-4.8}^{+1.0}$ & $2.4_{-0.2}^{+0.2}$ & Argument of pericenter (rad)\\
\hline
\multicolumn{6}{c}{\textit{Derived Transit Parameters}} \\
\hline
$i$ & $88.1_{-1.1}^{+1.2}$ & $88.8_{-0.3}^{+0.6}$ & $89.0_{-0.1}^{+0.1}$ & - & Inclination angle ($^{\circ}$) \\
$T_{\text{dur}}$ & $1.77_{-0.03}^{+0.03}$ & $2.77_{-0.04}^{+0.04}$ & $3.6_{-0.07}^{+0.08}$ & - & Transit duration\footnote{From first to last contact} (hrs)\\
\hline
\multicolumn{6}{c}{\textit{Derived Physical Parameters}} \\
\hline
$R_{p}$ & $3.02_{-0.08}^{+0.08}$ & $5.93_{-0.12}^{+0.11}$ & $6.18_{-0.15}^{+0.14}$ & - & Planet radius ($R_{\oplus}$)\\
$M_{p}$ & $11.00_{-1.27}^{+1.29}$ & $20.31_{-2.11}^{+2.13}$ & $38.15_{-3.22}^{+3.27}$ & $692_{-63}^{+66}$ & Planet mass\footnote{Minimum mass for TOI-4010 e} ($M_{\oplus}$) \\
$\rho_{p}$ & $2.2_{-0.3}^{+0.3}$ & $0.5_{-0.1}^{+0.1}$ & $0.9_{-0.1}^{+0.1}$ & - & Planet density (g cm$^{-3}$)\\
$a$ & $0.0229_{-0.0002}^{+0.0002}$ & $0.058_{-0.001}^{+0.001}$ & $0.113_{-0.001}^{+0.001}$ & $1.57_{-0.13}^{+0.12}$ & Semi-major axis (AU)\\
$S$ & $714_{-26}^{+28}$ & $112_{-4}^{+4}$ & $30_{-1}^{+1}$ & $0.15_{-0.02}^{+0.03}$ & Insolation flux ($S_{\oplus}$)\\
$T_{\text{eq}}$ & $1441_{-13}^{+14}$ & $907_{-8}^{+9}$ & $650_{-6}^{+6}$ & $174_{-6}^{+8}$ & Equilibrium temperature\footnote{Assuming albedo = 0} (K)\\
\hline
\multicolumn{6}{c}{\textit{Stellar Parameters}} \\
\hline
$M_\star$ ($M_\odot$) & \multicolumn{4}{c|}{$0.88_{-0.03}^{+0.03}$} & Stellar mass ($M_{\odot}$)\\
$R_\star$ ($R_\odot$) & \multicolumn{4}{c|}{$0.83_{-0.01}^{+0.01}$} & Stellar radius ($R_{\odot}$)\\
\hline
\multicolumn{6}{c}{\textit{Photometric Parameters}} \\
\hline
$\mu_{\text{FFI}}$ (ppm) & \multicolumn{4}{c|}{ $13_{-29}^{+29}$} & FFI flux offset (ppm)\\
$\mu_{\text{SPOC}}$ (ppm) & \multicolumn{4}{c|}{ $21_{-28}^{+27}$} & SPOC flux offset (ppm) \\
$\sigma_{\text{FFI}}$ (ppm) & \multicolumn{4}{c|}{ $48_{-33}^{+51}$} & FFI flux jitter (ppm)\\
$\sigma_{\text{SPOC}}$ (ppm) & \multicolumn{4}{c|}{ $70_{-49}^{+76}$} & SPOC flux jitter (ppm)\\
$q_{1,\text{TESS}}$ & \multicolumn{4}{c|}{ $0.29_{-0.18}^{+0.20}$} & Limb-darkening coefficient 1\footnote{\citet{exoplanet:kipping13} parameterization} \\
$q_{2,\text{TESS}}$ & \multicolumn{4}{c|}{ $0.28_{-0.36}^{+0.35}$} & Limb-darkening coefficient 2\\
\hline
\multicolumn{6}{c}{\textit{RV Parameters}}\\
\hline
$\mu_{\text{HARPS-N}}$ (m s$^{-1}$) & \multicolumn{4}{c|}{ $4781.6_{-6.6}^{+5.9}$} & HARPS-N RV offset (m s$^{-1}$) \\
$\sigma_{\text{HARPS-N}}$ (m s$^{-1}$) & \multicolumn{4}{c|}{ $3.7_{-0.6}^{+0.6}$} & HARPS-N RV jitter (m s$^{-1}$) \\
    \end{tabular}
\label{tab:fits}
\caption{Results of the four-planet joint fit of transit photometry and RV observations. The central values are the median of the posteriors, with uncertainties computed from the 16th and 84th percentiles.}  
\end{table*}

In summary, we find the following properties for the four planets of TOI-4010:

\begin{itemize}
    \item TOI-4010 b (TOI-4010.03) is a transiting inner sub-Neptune ($P = 1.3$ days, $R_{p} = 3.02_{-0.08}^{+0.08}~R_{\oplus}$, $M_{p} = 11.00_{-1.27}^{+1.29}~M_{\oplus}$, $e = 0.03_{-0.02}^{+0.03}$),
    \item TOI-4010 c (TOI-4010.01) is a transiting close-in sub-Saturn ($P = 5.4$ days, $R_{p} = 5.93_{-0.11}^{+0.13}~R_{\oplus}$, $M_{p} = 20.31_{-2.11}^{+2.13}~M_{\oplus}$, $e = 0.03_{-0.02}^{+0.03}$),
    \item TOI-4010 d (TOI-4010.02) is a more massive transiting close-in sub-Saturn ($P = 14.7$ days, $R_{p} = 6.18_{-0.15}^{+0.14}~R_{\oplus}$, $M_{p} = 38.15_{-3.22}^{+3.27}~M_{\oplus}$, $e = 0.07_{-0.03}^{+0.03}$), and
    \item TOI-4010 e is a long-period giant planet currently detected only in RV observations ($P \sim 762_{-90}^{+90}$ days, $M_{p}\sin{i} \sim 692_{-63}^{+66}~M_{\oplus}$, $e \sim 0.26_{-0.04}^{+0.04}$; preliminary solution due to insufficient RV baseline).
    \end{itemize}

As discussed in \S\ref{sec:RV}, we explored orbital periods for TOI-4010 e only up to 900 days, and chose a simple jitter term to capture RV noise rather than more complicated models, such as Gaussian Processes. These choices were due to the challenges of fitting a four-planet model with incomplete data and indications that TOI-4010 is a low-activity star. We do not expect our treatment of RV noise to affect the fitted RV semi-amplitudes and thus the derived masses of the planets, though it may affect our measurements of the orbital eccentricities \citep{Hara2019}. The RMS error of the RV residuals ($6.3~$m s$^{-1}$) is also slightly larger than the mean uncertainty of the RVs ($5~$m s$^{-1}$), indicating that there may be excess scatter due to activity. We leave fitting longer orbital periods and more complicated noise models to when more RV observations become available.

\section{Discussion}\label{sec:discussion}

The three-transiting-planet system bears a close resemblance to Kepler-18 \citep{Cochran2011} and K2-19 \citep{Armstrong2015}, which are the only other known systems with at least three planets at periods less than 15 days, two of which are larger than Neptune. All three systems feature two sub-Saturn-sized planets with a smaller inner planet. However, TOI-4010 is the only of these systems with an additional long-period giant companion, and the only configuration to lack strong resonances, which may be due to dynamical interaction with the moderately eccentric companion.

\begin{figure*}[t!]
    \centering
    \includegraphics[width=0.8\linewidth]{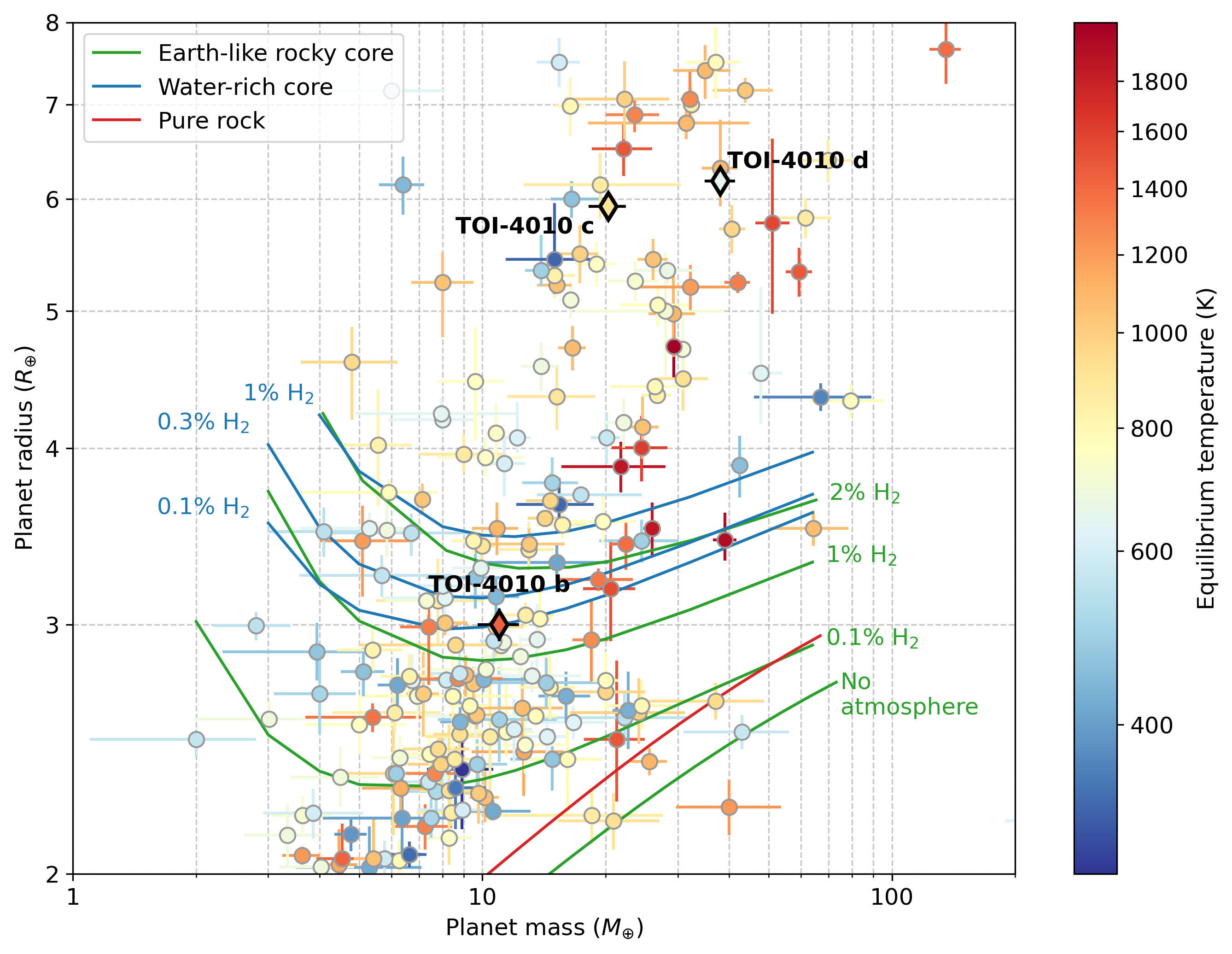}
    \caption{Mass-radius diagram for planets with density known to better than 50\% precision ($\sigma_{\rho_{p}}/\rho_{p} < 0.5$), plotted alongside the three transiting, short-period TOI-4010. Planets are coloured by equilibrium temperature, which we re-calculated assuming an albedo of 0 for uniformity. Mass-radius curves from \citet{Zeng2018} assuming 1000 K surface temperature for Earth-like rocky cores with no atmosphere or 0.1\%, 1\%, and 2\% H$_{2}$ envelopes by mass are overlaid in green, along with water-rich cores with 0.1\%, 0.3\%, and 1\% H$_{2}$ envelopes by mass (blue) and a pure rock composition (red).}
    \label{fig:mass_radius}
\end{figure*}

\begin{figure*}
    \centering
    \includegraphics[width=0.8\linewidth]{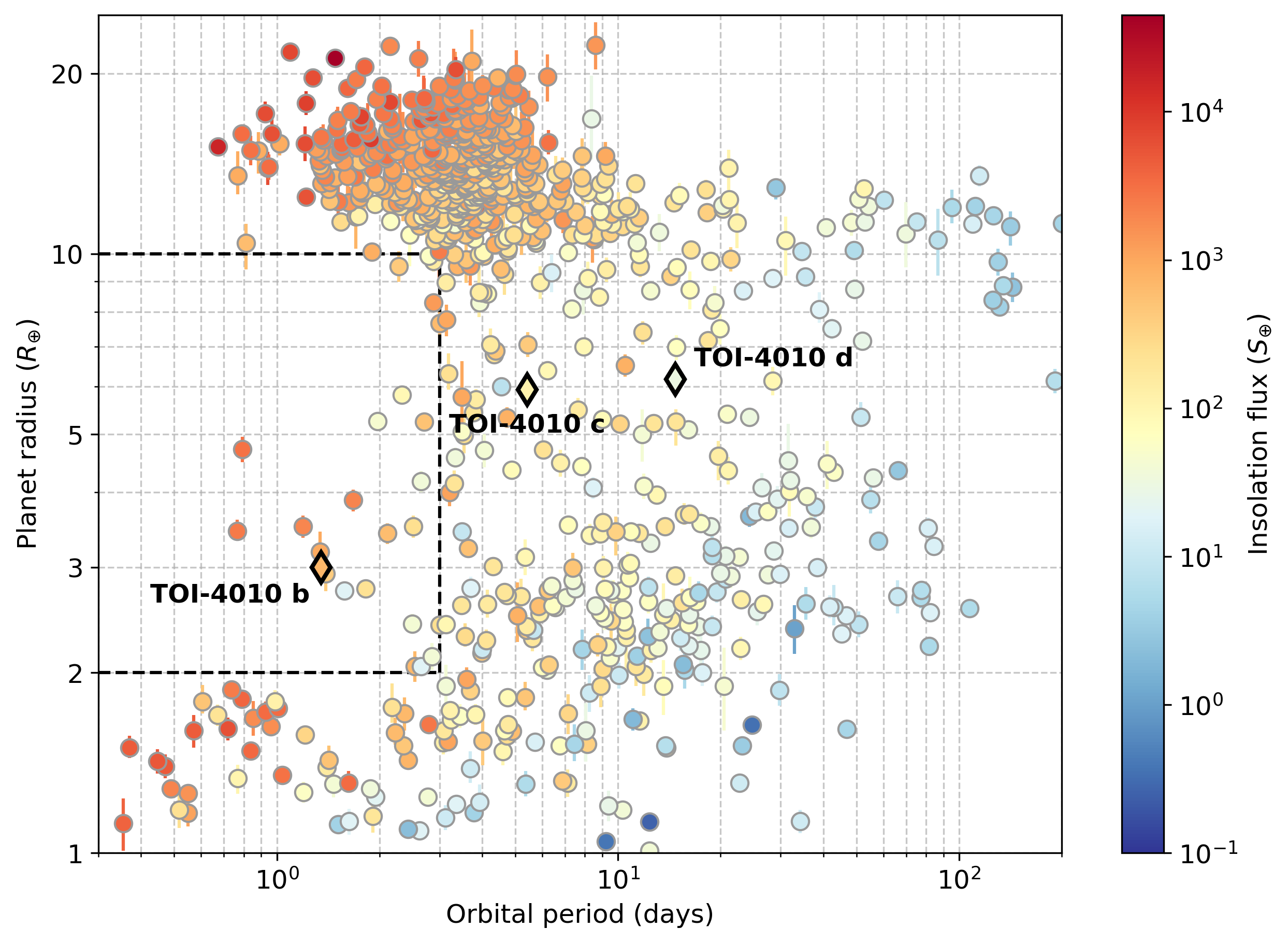}
    \caption{Period-radius diagram for the planets shown in Figure \ref{fig:mass_radius}, with planets coloured by insolation flux. TOI-4010 b lies in the sparsely populated region of short-period, intermediate-sized planets known as the hot Neptune desert. The edges of the desert are roughly indicated by the black dashed lines.}
    \label{fig:period_radius}
\end{figure*}

\begin{figure}
    \centering
    \includegraphics[width=0.8\linewidth]{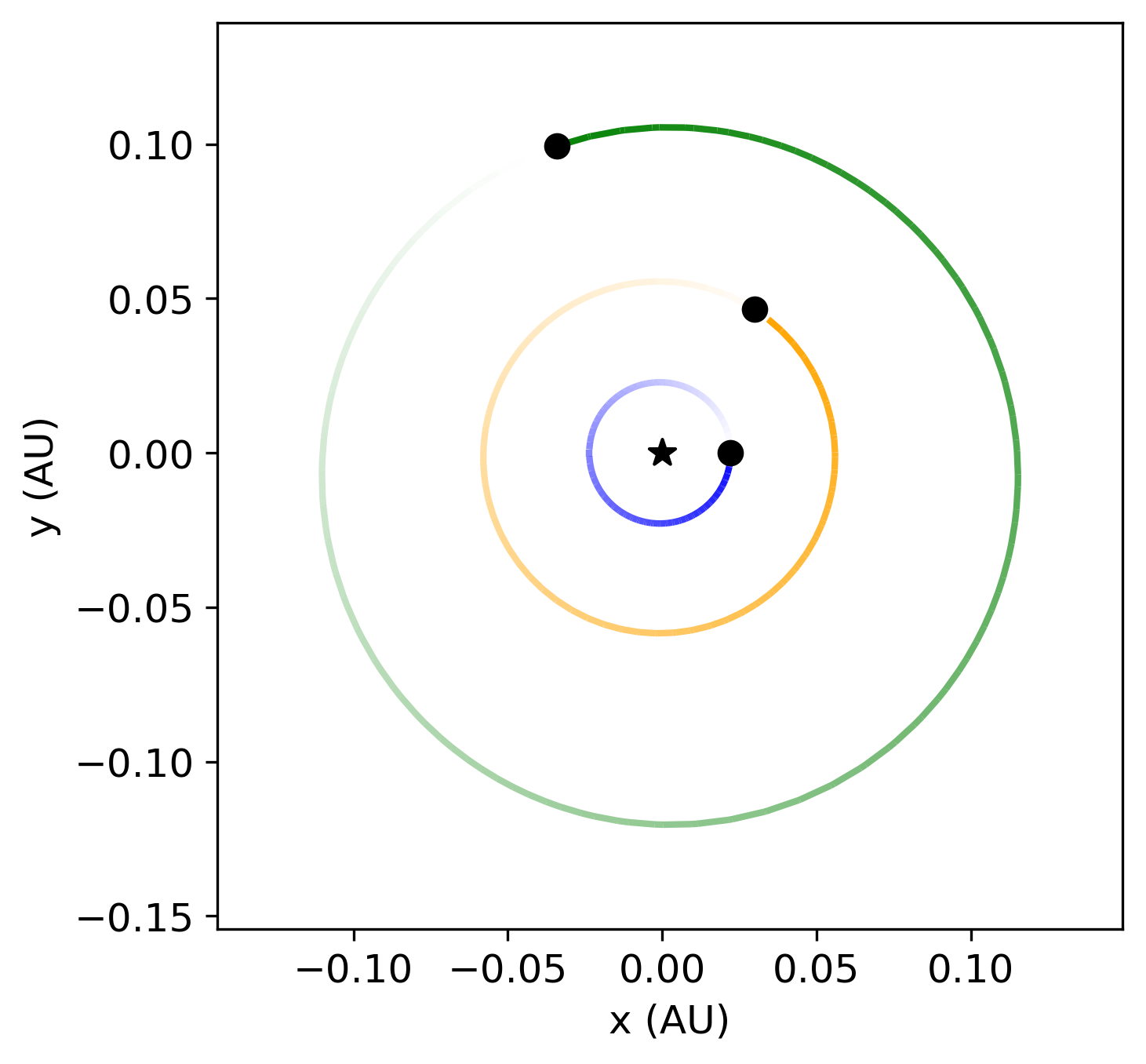}
    \caption{The orbits of the transiting planets of TOI-4010 assuming the median values provided in Table \ref{tab:fits}.}
    \label{fig:orbits}
\end{figure}

The transiting planets are plotted in mass-radius and period-radius diagrams in Figures \ref{fig:mass_radius} and \ref{fig:period_radius} alongside planets listed in the Planetary Systems Composite Data Table\footnote{Accessed 2023 March 13 at \url{https://exoplanetarchive.ipac.caltech.edu/cgi-bin/TblView/nph-tblView?app=ExoTbls&config=PSCompPars}} on the \citet{NEA}. TOI-4010 c and d land well within the large spread in typical masses ($M_{p}~\sim 5 - 80~M_{\oplus}$) for sub-Saturn-sized planets ($R_{p} = 4 - 8~R_{\oplus}$), while TOI-4010 b is slightly low-mass for its radius. The orbits of the planets are shown in Figure \ref{fig:orbits}.

\subsection{TOI-4010 b: A Sub-Neptune in the Hot Neptune Desert}

TOI-4010 b is an addition to the sparsely populated region of planets with $R_{p} \sim 2 - 10~R_{\oplus}$ and $P \lesssim 3$ days known as the hot Neptune (or sub-Jovian) desert \cite[e.g.][]{Szabo2011,Mazeh2016}. TOI-4010 b is especially unusual because it exists in a multi-planet system. Most planets in the desert lack planet companions \citep{Dong2018}, and TOI-4010 b is so far one of only two $P < 2$-day hot Neptunes with measured masses and radii that have known planet companions: the recently discovered TOI-969 b \cite[$P = 1.8$ days, $R_{p} = 2.8~R_{\oplus}$, $M_{p} = 9.1~M_{\oplus}$;][]{LilloBox2022} also has a companion, though at a significantly wider separation ($a \sim 2.5$ AU, $P \sim 1700$ days) than any of the other planets in the TOI-4010 system. As shown in Figure \ref{fig:period_radius}, TOI-4010 b's specific location in the desert appears to be in an oasis of hot Neptunes with $R_{p} \sim 3 - 5~R_{\oplus}$, above and below which there are almost no members of the desert with $P \lesssim 2$ days. It is unclear if this is a selection effect, such as a bias in selecting planets for RV follow-up, or an emerging feature of the hot Neptune population.

Both the upper and lower boundaries of the desert have been explained by tidal effects following high-eccentricity migration \citep{Matsakos2016, OwenLai2018}, though this mechanism may not be able to explain planets in or near the desert that are in multi-planet systems such as TOI-4010 b. The lower boundary could instead be sculpted by atmospheric mass-loss due to photoevaporation \citep{OwenLai2018}. However, TOI-4010 b still remains in the desert despite being highly irradiated (insolation flux $S = 714_{-26}^{+28} S_{\oplus}$). Furthermore, when compared to theoretical mass-radius curves from \citet{Zeng2018} for planets with $R_{p} = 2 - 4~R_{\oplus}$ at 1000 K surface temperature (a proxy for the $T_{\text{eq}} = 1441_{-13}^{+14}$ K measured for TOI-4010 b), we find that TOI-4010 b must still be maintaining an atmosphere. TOI-4010 b is most consistent with an Earth-like rocky core with $1 - 2\%$ H$_{2}$ envelope by mass, or water-rich core with $0.1 - 0.3\%$ H$_{2}$ envelope by mass (included in Figure \ref{fig:mass_radius}). The combination of hosting an atmosphere while still being highly irradiated may be explained by the fact that TOI-4010 b orbits a metal-rich star, and metal-rich atmospheres are less susceptible to photoevaporative mass-loss \citep{Owen2018, Wilson2022}.

\citet{Lundkvist2016} defined an alternative desert for hot super-Earths using insolation flux instead of period, with empirical bounds of $R_{p} = 2.2 - 3.8~R_{\oplus}$ and $S > 650 S_{\oplus}$. TOI-4010 b shares this desert with only five other confirmed planets with measured masses and radii: K2-66 b \citep{Sinukoff2017}, NGTS-4 b \citep{West2019}, TOI-132 b \citep{Diaz2020}, TOI-849 b \citep{Armstrong2020}, TOI-2196 b \citep{Persson2022}. All of these planets are remarkably different than TOI-4010 b, having two to five times the mass ($M_{p} = 21$ to $39~M_{\oplus}$) and up to three times the density ($\rho_{p} = 3.1$ to $7.6$~g~cm$^{-3}$) of TOI-4010 b ($M_{p} = 11.0 M_{\oplus}$, $\rho_{p} = 2.2$~g~cm$^{-3}$). The other members of the desert are also the only planets known in their systems, and their host stars are less metal-rich than TOI-4010 ([Fe/H]$~= -0.05$ to $0.19$ dex, compared to [Fe/H]$~= 0.37\pm0.07$ dex). These findings directly support the negative correlation between planet density and host-star metallicity found for sub-Neptunes by \citet{Wilson2022}, further suggesting that high metallicity shaped their histories. TOI-4010 b will be a unique target for future studies on highly irradiated sub-Neptunes.

\subsection{TOI-4010 c and d: A Pair of Similar Sub-Saturns}

\begin{figure}[t!]
    \centering
    \includegraphics[width=\linewidth]{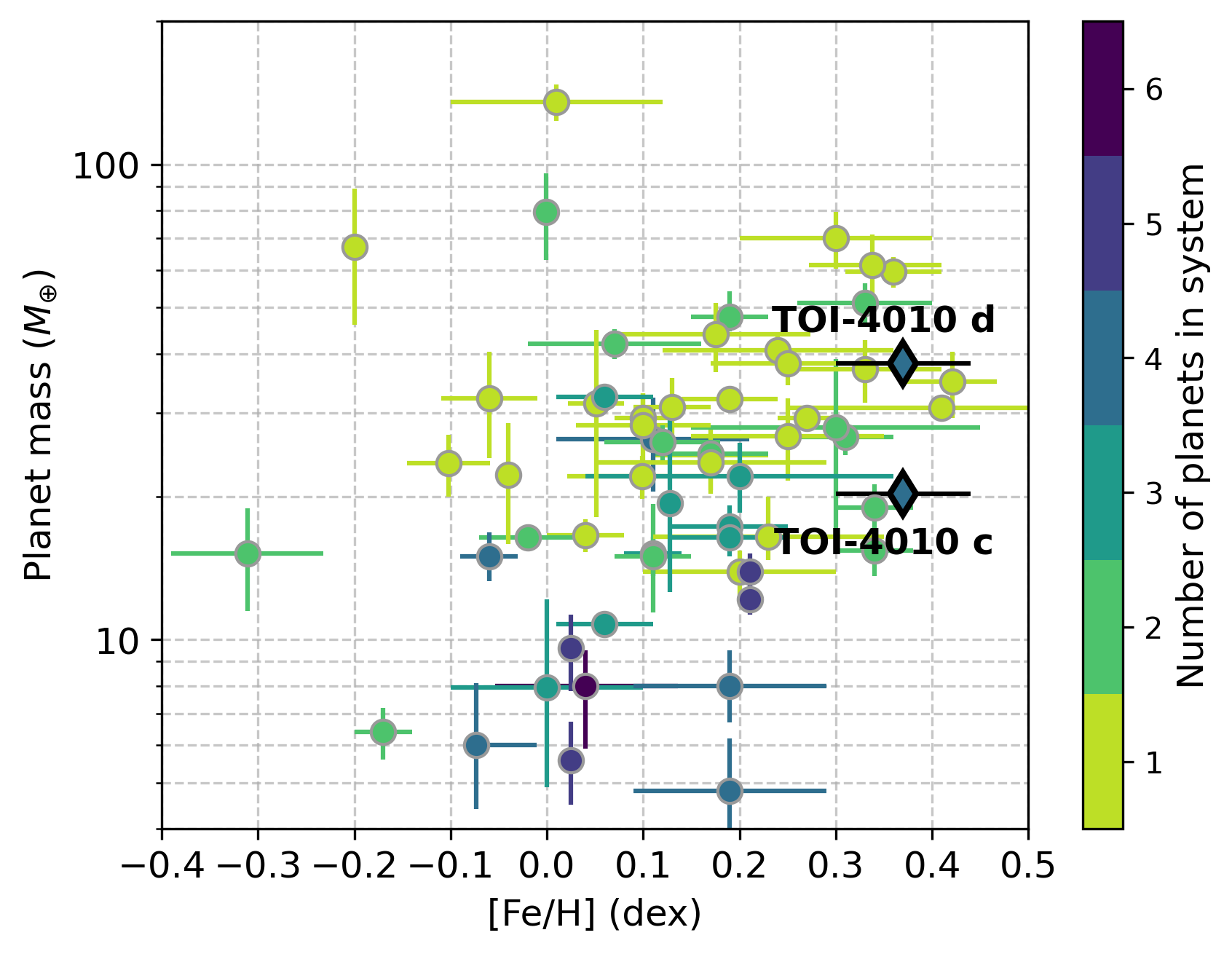}
    \caption{Mass-metallicity diagram for sub-Saturn-sized planets ($R_{p} = 4 - 8~R_{\oplus}$) from Figure \ref{fig:mass_radius}, including TOI-4010 c and d. This population shows a tentative positive correlation between mass and host star metallicity. TOI-4010 d is one of the most massive sub-Saturns discovered in a multi-planet system. TOI-4010 follows the general trend of more massive sub-Saturns orbiting higher-metallicity stars.}
    \label{fig:toib_c}
\end{figure}

TOI-4010 hosts two similarly sized sub-Saturns, one of which (TOI-4010 d) is among the most massive sub-Saturns discovered in a multi-planet system to date (Figure \ref{fig:toib_c}). When compared to other systems with multiple short-period sub-Saturns, TOI-4010 stands out further: the combined mass of TOI-4010 c and d is $\sim58~M_{\oplus}$, greater than the combined masses of the two sub-Saturns orbiting Kepler-18 \cite[$\sim34~M_{\oplus}$;][]{Cochran2011} and K2-19 \cite[$\sim43~M_{\oplus}$;][]{Petigura2020}.

Giant planets are well-established to be more abundant around more metal-rich stars \citep[e.g.][]{Fischer2005,Johnson2010,Petigura2018}, and \citet{Petigura2017} found a positive correlation between sub-Saturn mass and host star metallicity. These correlations have been interpreted as support for core accretion theory, given that high metallicity disks should have more solid materials available to form more massive cores. TOI-4010 is fittingly one of the most metal-rich sub-Saturn hosts known so far, with [Fe/H] = $0.37\pm0.07$ dex, and only Kepler-27 \citep{Steffen2012} is a more metal-rich star hosting a multi-planet system with a sub-Saturn. However, as shown in Figure \ref{fig:toib_c}, a more updated sub-Saturn population with well-measured densities (precision better than 50\%) reveals a weaker correlation between sub-Saturn mass and host star metallicity compared to \citet{Petigura2017}. We find a Spearman's rank correlation coefficient of $r_{s} = +0.19$, indicating a weak positive correlation. This correlation is slightly stronger among sub-Saturns in multi-planet systems ($r_{s} = +0.33$), and moderate among sub-Saturns with $M_{p} < 60 M_{\oplus}$ ($r_{s} = +0.37$). While the relatively large masses of TOI-4010 b and c follow this general trend, additional planets will be needed to better characterize the strength of the correlation.

\subsection{TOI-4010 e: A Long-Period Giant Planet Companion}

Our RV observations reveal an outer companion with a possible orbital period of $P = 762_{-90}^{+90}$ days and minimum mass of $M_{p}\sin{i} = 692_{-63}^{+66}~M_{\oplus}$ ($2.2_{-0.2}^{+0.2} M_{Jup}$) consistent with at least a super-Jupiter. At this period, the planet receives an insolation flux of $S = 0.15_{-0.02}^{+0.03} S_{\oplus}$, placing it just outside of its host star's habitable zone \citep{Kopparapu2013}.

Based on the fitted next transit time of $3218_{-86}^{+86}$ TBJD (BJD - 2457000) from the RV observations, no transits of TOI-4010 e were expected to occur during any of the TESS sectors over which TOI-4010 was observed. We cannot yet know whether or not TOI-4010 e has a transiting orbital geometry. Without a transit, we can not know the radius, orbital inclination, and true mass of TOI-4010 e. Nevertheless, we will continue to monitor TOI-4010 with RVs to improve orbital phase coverage and better constrain its period and eccentricity.

We explored what constraints Gaia absolute astrometry can provide on the true mass of TOI-4010 e. The Gaia DR3 archive entry for TOI-4010 reports a Renormalized Unit Weight Error (RUWE) value of 0.84. A RUWE $>1.4$ \citep{Lindegren2018,Lindegren2021} is typically adopted as a threshold to indicate significant departure from a single-star model and likely evidence of orbital motion in Gaia astrometry. The star thus appears well-described as a single star. We then performed numerical simulations following the approach by e.g. \citet{Belokurov2020} to constrain the regime of companions masses and orbital separations that would produce excess astrometric residuals with respect to a single-star model, and therefore a RUWE value larger than the one reported in the Gaia DR3 archive. At the currently assumed separation of TOI-4010 e ($a = 1.57_{-0.12}^{+0.13}$ AU), its true mass would need to be $>6$ M$_\mathrm{Jup}$ in order to record a larger RUWE with 99\% probability. Gaia DR3 astrometry thus allows us to rule out quasi-face-on orbital configurations ($i\lesssim20^\circ$) for TOI-4010 e. Significantly improved true mass constraints even in the case of a close-to-coplanar orbit for this massive companion are expected to come from Gaia DR4, slated to be made public in late 2025.

\subsection{Dynamical Stability}

TOI-4010 b and c are within 2\% of a 4:1 (third order) mean motion resonance (period ratio 4.0158), but we do not expect there to be significant transit timing variation (TTVs) because no adjacent planet pairs are near strong mean motion resonances. We ran the three transiting, short-period planets through 1000 simulations using \texttt{ttvfast-python}, a Python wrapper for the TTVFast symplectic integrator for TTV computation \citep{Deck2014}. We ran the integrator for 954 days (the total timespan of our TESS photometry) with a time step of 0.067 days (1/20th the period of TOI-4010 b). The median largest TTV amplitude for any planet in the system was less than 2 minutes.

We used the SPOCK stability classifier \citep{Tamayo2020} to assess the likely long-term stability of the system. SPOCK is an XGBoost-based machine learning classifier trained with 100,000 compact 3-planet systems to quickly predict the probability that a system will remain dynamically stable for $10^9$ orbits of the innermost planet, and was directly demonstrated in \citet{Tamayo2020} to give reliable results for systems of up to five planets.  We ran the SPOCK classifier on the posterior from the joint fit under two separate assumptions for the unmeasured inclination of TOI-4010 e: one in which the mutual inclination with the inner system was 0, and another where the mutual inclination was drawn from a uniform distribution between -10$^\circ$ and 10$^\circ$. For each assumption, we computed the stability probability for every draw from the posterior. Both cases showed a high level of stability for the system (with the majority of draws having stability probabilities upwards of 0.84. We find that the least dynamically stable configurations have larger eccentricities for TOI-4010 c and d or a higher mutual inclination between those two planets. 

We also used full N-body integrations to constrain the feasible parameter space for the orbital inclination of TOI-4010 e. We first randomly selected a draw from the posterior and chose the inclination of TOI-4010 e from a grid spanning a mutual inclination with the inner system between -85$^\circ$ to 85$^\circ$. We initialized each simulation with the drawn parameters, chosen inclination for $i_{e}$, and $M_{e}$ computed from the drawn value of $M_{e}\sin{i_{e}}$. We then integrated each set of initial conditions for 1 Myr in \texttt{rebound} \citep{rebound, reboundx, reboundias15} using the MERCURIUS integrator and the additional general relativistic correction implemented in the REBOUNDx gr effect. Almost all of these simulations were dynamically stable for the full integration time, but the integrations with 85$^\circ$ mutual inclination went dynamically unstable (signified by destruction of the inner system) within a few thousand years. This result suggests that we can use the stability of the inner system to constrain the mutual inclination of TOI-4010 e to be less than 85$^\circ$ relative to the plane of transiting planets, leaving a very large range of allowable parameter space for the orbital inclination of TOI-4010 e where its presence will not disrupt the inner system.

\subsection{Follow-Up Feasibility}

We computed the Transmission Spectroscopy Metric \cite[TSM;][]{Kempton2018} for each transiting planet using their derived planet masses. The TSM measures how promising planets are for atmospheric characterization studies. We find TSM = 49, 115, and 50 for TOI-4010 b, c, and d, respectively. When compared to other planets with densities measured to better than 50\% precision, TOI-4010 b has a higher TSM than 72\% of planets with $R_{p} < 4~R_{\oplus}$, and TOI-4010 c and d have higher TSM than 88\% and 56\% of other sub-Saturns in multi-planet systems, respectively. Importantly, TOI-4010 has multiple promising planets, which can enable atmospheric comparative planetology studies.

\section{Summary}

We have presented the confirmation of three transiting exoplanets and the discovery of a fourth planet orbiting TOI-4010, a metal-rich ([Fe/H] $=0.37\pm0.07$ dex) K dwarf observed by TESS in Sectors 24, 25, 52, and 58:

\begin{itemize}
    \item TOI-4010 b (TOI-4010.03; $P = 1.3$ days, $R_{p} = 3.02_{-0.08}^{+0.08}~R_{\oplus}$, $M_{p} = 11.00_{-1.27}^{+1.28}~M_{\oplus}$) is an addition to the census of planets in the hot Neptune desert, specifically occupying an oasis of hot Neptunes with $R_{p} \sim 3 - 5~R_{\oplus}$, which may be an emerging feature of the desert. Despite being highly irradiated, TOI-4010 b may still hold on to an atmosphere, likely due to having a metal-rich atmosphere that resists photoevaporative mass-loss. Compared to similarly irradiated planets near its size with both measured mass and radius, it is significantly less massive and dense, and the only one in a multi-planet system.
    \item TOI-4010 c (TOI-4010.01; $P = 5.4$ days, $R_{p} = 5.93_{-0.11}^{+0.13}~R_{\oplus}$, $M_{p} = 20.31_{-2.11}^{+2.13}~M_{\oplus}$) is a close-in sub-Saturn near the 4:1 mean-motion resonance with TOI-4010 b.
    \item TOI-4010 d (TOI-4010.02; $P = 14.7$ days, $R_{p} = 6.18_{-0.15}^{+0.14}~R_{\oplus}$, $M_{p} = 38.15_{-3.22}^{+3.27}~M_{\oplus}$) is a more massive close-in sub-Saturn, and is the most massive sub-Saturn-sized planet discovered in a system with at least three planets.
    \item TOI-4010 e ($P \sim 762_{-90}^{+90}$ days, $M_{p}\sin{i} \sim 692_{-63}^{+66}~M_{\oplus}$) is a moderately eccentric ($e \sim 0.26_{-0.04}^{+0.04}$), long-period giant planet companion detected in our RV observations. These are preliminary physical and orbital properties based on exploring possible orbits with periods up to 900 days due to a lack of sufficient RV baseline. We will continue to obtain observations to better constrain the orbit and determine whether or not the period may be longer.
\end{itemize}TOI-4010 was followed-up with ground-based photometry, high-resolution imaging, and reconnaissance spectroscopy, and all three transiting planets were confirmed with precise RV observations from HARPS-N. This is only the sixth discovered system with two sub-Saturn-sized planets with $P < 20$ days, and the only known system containing three planets with $R_{p} > 3 R_{\oplus}$ in $P < 15$-day orbits. TOI-4010 c and d are the most massive planets discovered among other rare multi-planet systems with multiple sub-Saturn-sized planets. Sub-Saturns as a population exhibit a large diversity in planet mass at a given size, and two such planets in the same system present an opportunity to study their properties with other variables (e.g. host star age, metallicity) held constant. The TOI-4010 system currently has no known analogues, making it an intriguing target for future work on planet formation and evolution theory.

\section{Acknowledgements}

We thank the referee for their comments and suggestions, which improved the quality of the manuscript.

% TESS
This paper includes data collected by the TESS mission, which are publicly available from the Mikulski Archive for Space Telescopes (MAST). Funding for the TESS mission is provided by NASA's Science Mission directorate. We acknowledge the use of public TESS data from pipelines at the TESS Science Office and at the TESS Science Processing Operations Center. Resources supporting this work were provided by the NASA High-End Computing (HEC) Program through the NASA Advanced Supercomputing (NAS) Division at Ames Research Center for the production of the SPOC data products.

% HARPS-N
The HARPS-N project was funded by the Prodex Program of the Swiss Space Office (SSO), the Harvard University Origin of Life Initiative (HUOLI), the Scottish Universities Physics Alliance (SUPA), the University of Geneva, the Smithsonian Astrophysical Observatory (SAO), and the Italian National Astrophysical Institute (INAF), University of St. Andrews, Queen’s University Belfast and University of Edinburgh.

% Gaia
This work has made use of data from the European Space Agency (ESA) mission Gaia (\url{https://www.cosmos.esa.int/gaia}), processed by the Gaia Data Processing and Analysis Consortium (\url{DPAC, https://www.cosmos.esa.int/web/gaia/dpac/consortium}). Funding for the DPAC has been provided by national institutions, in particular the institutions participating in the Gaia Multilateral Agreement.

% ExoFOP
This research has made use of the Exoplanet Follow-up Observation Program website, which is operated by the California Institute of Technology, under contract with the National Aeronautics and Space Administration under the Exoplanet Exploration Program. 

% NASA Exoplanet Archive
This research has made use of the NASA Exoplanet Archive, which is operated by the California Institute of Technology, under contract with the National Aeronautics and Space Administration under the Exoplanet Exploration Program. 

% LCO
This work makes use of observations from the LCOGT network. This paper is based on observations made with the MuSCAT3 instrument, developed by the Astrobiology Center and under financial supports by JSPS KAKENHI (JP18H05439) and JST PRESTO (JPMJPR1775), at Faulkes Telescope North on Maui, HI, operated by the Las Cumbres Observatory.

% Misc.
This project has received funding from the European Research Council (ERC) under the European Union’s Horizon 2020 research and innovation programme (grant agreement SCORE No 851555). This work has been carried out within the framework of the NCCR PlanetS supported by the Swiss National Science Foundation under grants 51NF40\textunderscore 182901 and 51NF40\textunderscore 205606. R.D.H. is funded by the UK Science and Technology Facilities Council (STFC)'s Ernest Rutherford Fellowship (grant number ST/V004735/1). T.~Dupuy acknowledges support from UKRI STFC AGP grant ST/W001209/1. TGW acknowledges support from STFC consolidated grant numbers
ST/R000824/1 and ST/V000861/1, and UKSA grant number ST/R003203/1. A.S. and M.Pi. acknowledge financial contribution from the agreement ASI-INAF n.2018-16-HH.0. M.Pi. acknowledge financial contribution from the PRIN-INAF 2019 "Planetary systems at young ages (PLATEA)." B.S.S. and M.V.G. acknowledge the support of Ministry of Science and Higher Educationof the Russian Federation under the grant 075-15-2020-780 (N13.1902.21.0039). D.H. acknowledges support from the Alfred P. Sloan Foundation, the National Aeronautics and Space Administration (80NSSC21K0652), and the National Science Foundation (AST-2009828). J.B. acknowledges support from the National Science Foundation (AST-2009828). For the purpose of open access, the author has applied a Creative Commons Attribution (CC BY) licence to any Author Accepted Manuscript version arising from this submission.

\facilities{Exoplanet Archive, FLWO:1.2m (KeplerCam), FLWO:1.5m (TRES), \gaia\, Keck:II (NIRC2), LCOGT, TESS, TNG  (HARPS-N)}

\software{\texttt{emcee} \citep{ForemanMackey2013}, \texttt{exoplanet} \citep{exoplanet:joss,
exoplanet:zenodo, exoplanet:agol20,
exoplanet:arviz, exoplanet:astropy13, exoplanet:astropy18, exoplanet:kipping13,
exoplanet:luger18, exoplanet:pymc3, exoplanet:theano}, \texttt{isochrones} \citep{Morton2015}, \texttt{matplotlib} \citep{Hunter2007}, \texttt{numpy} \citep{Harris2020}, \texttt{pandas} \citep{reback2020pandas, mckinney-proc-scipy-2010}, \texttt{PYMC3} \citep{exoplanet:pymc3}, \texttt{scipy} \citep{Virtanen2020}}

\appendix

\renewcommand{\thefigure}{A\arabic{figure}}
\renewcommand{\thetable}{A\arabic{table}}

\setcounter{table}{0}
\setcounter{figure}{0}

\begin{figure}
    \centering
    \includegraphics[width=\linewidth]{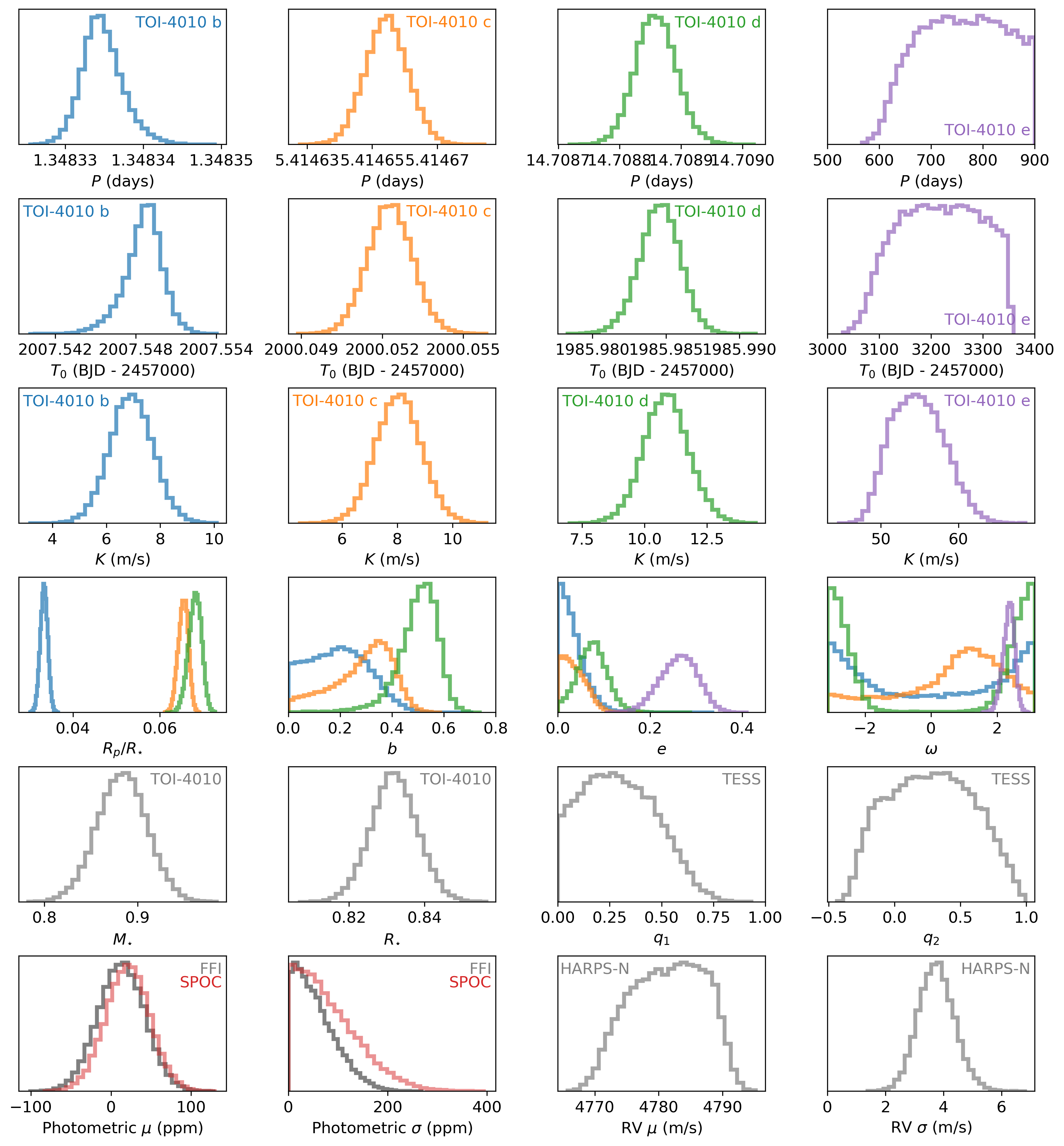}
    \caption{Histograms of the distributions of all fit parameters from the four-planet, joint transit + RV fit. The parameters fitted for TOI-4010 b, c, d, and e are shown in blue, orange, green, and purple, respectively.}
    \label{fig:posteriors}
\end{figure}

The priors used for our joint transit + RV model fit (\S\ref{sec:joint}) are listed in Table \ref{tab:priors}, while the final distributions of all fit parameters are shown in Figure \ref{fig:posteriors}.

\begin{table*}[h!]
    \centering
    \begin{tabular}{l|l|l}
    \hline\hline
    Parameter & Prior & Description\\
\hline\hline
\multicolumn{3}{c}{\textit{Planet Parameters: TOI-4010 b}} \\
\hline
$P$ & $\mathcal{N}(1.3483,0.1348)$\footnote{$\mathcal{N}(\mu,\sigma)$ denotes a normal prior with mean $\mu$ and standard deviation $\sigma$} & Orbital period (days) \\
$T_{0}$ & $\mathcal{N}(2007.5486,0.1348)$ & Transit epoch (BJD $-$ 2457000) \\
$K$ & $\mathcal{U}(0,100)$ & RV semi-amplitude (m~s$^{-1}$) \\
$R_p/R_\star$ & $\mathcal{U}(0,0.5)$ & Planet-to-star radius ratio \\
$b$ & $\mathcal{U}(0,1)$ & Impact parameter \\
$e$ & \citet{VanEylen2019} & Orbital eccentricity \\
$\omega$ & $\mathcal{U}(-\pi, \pi)$ & Argument of pericenter (rad)\\
\hline
\multicolumn{3}{c}{\textit{Planet Parameters: TOI-4010 c}} \\
\hline
$P$ & $\mathcal{N}(5.4146,0.5415)$ & Orbital period (days) \\
$T_{0}$ & $\mathcal{N}(2000.0520,0.5415)$ & Transit epoch (BJD $-$ 2457000) \\
$K$ & $\mathcal{U}(0,100)$ & RV semi-amplitude (m~s$^{-1}$) \\
$R_p/R_\star$ & $\mathcal{U}(0,0.5)$ & Planet-to-star radius ratio \\
$b$ & $\mathcal{U}(0,1)$ & Impact parameter \\
$e$ & \citet{VanEylen2019} & Orbital eccentricity \\
$\omega$ & $\mathcal{U}(-\pi, \pi)$ & Argument of pericenter (rad)\\
\hline
\multicolumn{3}{c}{\textit{Planet Parameters: TOI-4010 d}} \\
\hline
$P$ & $\mathcal{N}(14.7088,1.4709)$ & Orbital period (days) \\
$T_{0}$ & $\mathcal{N}(1985.9844,1.4709)$ & Transit epoch (BJD $-$ 2457000) \\
$K$ & $\mathcal{U}(0,100)$ & RV semi-amplitude (m~s$^{-1}$) \\
$R_p/R_\star$ & $\mathcal{U}(0,0.5)$ & Planet-to-star radius ratio \\
$b$ & $\mathcal{U}(0,1)$ & Impact parameter \\
$e$ & \citet{VanEylen2019} & Orbital eccentricity \\
$\omega$ & $\mathcal{U}(-\pi, \pi)$ & Argument of pericenter (rad)\\
\hline
\multicolumn{3}{c}{\textit{Planet Parameters: TOI-4010 e}} \\
\hline
$P$ & $\mathcal{U}(500,900)$ & Orbital period (days) \\
$T_{0}$ & $\mathcal{U}(2700,3400)$ & Transit epoch (BJD $-$ 2457000) \\
$K$ & $\mathcal{U}(0,100)$ & RV semi-amplitude (m~s$^{-1}$) \\
$e$ & \citet{Kipping2013a} & Orbital eccentricity \\
$\omega$ & $\mathcal{U}(-\pi, \pi)$ & Argument of pericenter (rad)\\
\hline
\multicolumn{3}{c}{\textit{Stellar Parameters}} \\
\hline
$M_{\star}$ & $\mathcal{N}(0.88,0.03)$ & Stellar mass ($M_{\odot}$)\\
$R_{\star}$ & $\mathcal{N}(0.83,0.01)$ & Stellar radius ($R_{\odot}$) \\
\hline
\multicolumn{3}{c}{\textit{Photometric Parameters}} \\
\hline
$\mu_{\text{FFI}}$ & $\mathcal{N}(0,100)$ & FFI flux offset (ppm) \\
$\mu_{\text{SPOC}}$ & $\mathcal{N}(0,100)$ & SPOC flux offset (ppm) \\
$\sigma_{\text{FFI}}$ & $\mathcal{H}(1000)$\footnote{$\mathcal{H}(\sigma)$ denotes a half-normal prior with standard deviation $\sigma$} & FFI flux jitter (ppm) \\
$\sigma_{\text{SPOC}}$ & $\mathcal{H}(1000)$ & SPOC flux jitter (ppm) \\
$q_{1}$ & $\mathcal{U}(-1,1)$ & Limb-darkening coefficient 1 \\
$q_{2}$ & $\mathcal{U}(-1,1)$ & Limb-darkening coefficient 2\\
\hline
\multicolumn{3}{c}{\textit{RV Parameters}} \\
\hline
$\mu_{\text{HARPS-N}}$ & $\mathcal{U}(4692,4799)$ & HARPS-N RV offset (m~s$^{-1}$) \\
$\sigma_{\text{HARPS-N}}$ & $\mathcal{H}(10)$ & HARPS-N RV jitter (m~s$^{-1}$) \\
    \end{tabular}
    \caption{Priors placed on all model parameters for the joint fit of transit photometry and RV observations.}
    \label{tab:priors}
\end{table*}

\bibliography{sample631}{}
\bibliographystyle{aasjournal}

\end{document}